\documentclass[11pt,english]{article}
\usepackage[T1]{fontenc}
\usepackage[latin9]{inputenc}
\usepackage{geometry}
\usepackage{color}
\usepackage{amsmath}
\usepackage{amssymb}
\usepackage{graphicx}
\usepackage{setspace}
\usepackage[authoryear]{natbib}
\usepackage{hyperref}
\makeatletter

\providecommand{\tabularnewline}{\\}

\@ifundefined{date}{}{\date{}}

\bibpunct{(}{)}{,}{autheryear}{,}{,}

\newtheorem{assumption}{Assumption}

\newtheorem{theorem}{Theorem}

\allowdisplaybreaks

\newtheorem{corollary}{Corollary}

\newtheorem{proposition}{Proposition}
\newtheorem{definition}{Definition}
\newtheorem{lemma}{Lemma}

\newtheorem{remark}{Remark}

\newcommand*{\indep}{%
  \mathbin{%
    \mathpalette{\@indep}{}%
  }%
}
\newcommand*{\nindep}{%
  \mathbin{
    \mathpalette{\@indep}{\not}
  }%
}
\newcommand*{\@indep}[2]{%
  \sbox0{$#1\perp\m@th$}
  \sbox2{$#1=$}
  \sbox4{$#1\vcenter{}$}
  \rlap{\copy0}
  \dimen@=\dimexpr\ht2-\ht4-.2pt\relax
  \kern\dimen@
  {#2}%
  \kern\dimen@
  \copy0 
} 

\newcommand{\pr}{P} 
\newcommand{\cA}{\mathcal{A}}
\newcommand{\cB}{\mathcal{B}}
\newcommand{\cC}{\mathcal{C}}
\newcommand{\F}{\mathcal{F}}
\newcommand{\var}{\mathrm{var}}
\newcommand{\cov}{\mathrm{cov}}
\newcommand{\avar}{\mathrm{a.var}}
\newcommand{\acov}{\mathrm{a.cov}}
\newcommand{\corr}{\mathrm{corr}}

\newcommand{\T}{\mathrm{\scriptscriptstyle T}}

\newcommand{\reg}{\mathrm{reg}}

\newcommand{\I}{\mathrm{I}}
\newcommand{\II}{\mathrm{II}}
\newcommand{\III}{\mathrm{III}}

\newcommand{\N}{\mathcal{N}}

\newcommand{\as}{\mathrm{a.s.}}
\usepackage{fancyhdr}

\usepackage{babel}

\usepackage{babel}

\usepackage{babel}

\makeatother

\usepackage{babel}
\begin{document}
\title{{\huge\textbf{Two-phase rejective sampling and its asymptotic properties}}}
\author{Shu Yang\thanks{Department of Statistics, North Carolina State University, North Carolina
27695, U.S.A. Email: syang24@ncsu.edu}$\ \ $and Peng Ding\thanks{Department of Statistics, University of California, Berkeley, California
94720, U.S.A. Email: pengdingpku@berkeley.edu}}
\maketitle
\begin{abstract}
Rejective sampling improves design and estimation efficiency of single-phase
sampling when auxiliary information in a finite population is available.
When such auxiliary information is unavailable, we propose to use
two-phase rejective sampling (TPRS), which involves measuring auxiliary
variables for the sample of units in the first phase, followed by
the implementation of rejective sampling for the outcome in the second
phase. We explore the asymptotic design properties of double expansion
and regression estimators under TPRS. We show that TPRS enhances the
efficiency of the double expansion estimator, rendering it comparable
to a regression estimator. We further refine the design to accommodate
varying importance of covariates and extend it to multi-phase sampling.
\textcolor{black}{We start with the theory for the population mean
and then extend the theory to parameters defined by general estimating
equations.} Our asymptotic results for TPRS immediately cover the
existing single-phase rejective sampling, under which the asymptotic
theory has not been fully established. 

\bigskip{}
\end{abstract}

\section{Introduction\protect\label{sec:Introduction}}

Two-phase sampling, also known as double sampling, is a cost-effective
method in large surveys, initially using auxiliary variables for broad
measurement in the first phase, followed by targeted measurement of
primary study variables in the second phase. Pioneered by \citet{neyman1938contribution}
and further developed by \citet{cochran2007sampling} and \citet{legg2009two},
two-phase sampling integrates auxiliary variables in both the design
and analysis stages. In the design stage, stratified sampling is often
used to leverage discrete auxiliary variables from the first phase
to inform the selection of the second-phase sample. This strategy
enhances the efficiency of estimators like weighted expansion estimators
\citep{oh1983weighting} over simple random sampling of the same size.
In the analysis stage, auxiliary variables can be used to enhance
the efficiency of estimation, for example, through regression adjustment.
Regression estimators work for both discrete and continuous variables,
providing gains in efficiency over traditional approaches without
auxiliary variables.

Regression estimators, however, have the practical drawback of having
potential negative weights. Methods such as rejective sampling\textcolor{black}{{}
(\citealp{hajek1964asymptotic}; \citealp{hajek1981sampling})} and
balanced sampling (\citealp{yates1946review}; \citealp{valliant2000finite};
\citealp{deville2004efficient,deville2005variance}; \citealp{tille2005optimal})
have been developed to mitigate the problem of negative weights associated
with regression estimators.\textcolor{blue}{{} }\textcolor{black}{Rejective
sampling involves selecting a sample through a basic sampling design,
which is then rejected if the difference between the sample mean and
the population mean of an auxiliary vector exceeds a specified threshold.
Extensive analyses of such samples are provided by \citet{hajek1964asymptotic}
and \citet{hajek1981sampling}. \citet{herson1976investigation} extends
this by describing a rejection procedure where the rejection region
is defined as a cube, and introduces the concept of a balanced sample---later
referred to as partially balanced or restricted samples by others.
Balanced sampling ensures that the sample mean closely matches the
population mean of auxiliary variables, retaining the optimality of
the ratio estimator under many polynomial regression models \citep{royall1973robust}.
\citet{valliant2000finite} further explore the use of balanced or
partially balanced samples with model-based estimators. \citet{deville2004efficient}
introduce the cube method for creating such balanced samples, even
with unequal probabilities and multiple auxiliary variables. They
also establish that a conditional Poisson design with appropriate
inclusion probabilities can be used to create a balanced sample \citep{deville2005variance}.
\citet{fuller2009some} demonstrates that the mean and variance of
the regression estimator are asymptotically equivalent for both rejective
and original samples, and that the variance estimator for the original
sample is appropriate for the rejective sample. Empirical comparisons
of the cube method and rejective sampling are provided by \citet{Legg2010}.
\citet{fuller2017bootstrap} introduce a bootstrap variance estimation
method for single-phase rejective sampling. \citet{yang2023rejective}
combine rejective sampling and rerandomization in experiments to improve
both external and internal validity. However, most of the existing
design methods are limited to single-phase sampling and require auxiliary
variable data for the entire population.}

We introduce two-phase rejective sampling (TPRS) and explore its asymptotic
design properties with commonly-used estimators, namely weighted expansion
and regression estimators.\textcolor{blue}{{} }\textcolor{black}{In
design-based inference, the characteristics of finite populations
are fixed, and the randomness arises from the sampling process.}\textcolor{blue}{{}
}TPRS allows for using both continuous and discrete auxiliary variables
in the design stage, relaxing the requirement of observing auxiliary
variables in the whole finite population typically associated with
single-phase sampling. TPRS offers several practical benefits: it
ensures an unbiased sample of the target population, reduces the variance
of the population mean estimator for covariates, prevents the selection
of samples with extreme auxiliary variable values, and reduces the
likelihood of negative weights in regression estimators. Additionally,
\textcolor{black}{TPRS enhances the efficiency of double-expansion
estimators across multiple outcomes to match the performance of regression
estimators, without the need of multiple model fitting for outcomes.
We present the first derivation of the asymptotic distribution result
under TPRS, complementing \citet{fuller2009some}'s work, which established
the consistency and asymptotic variance of the regression estimator
without deriving the asymptotic distribution under single-phase rejective
sampling. Furthermore, we refine the TPRS design to account for varying
importance of covariates and extend it to multi-phase sampling. Under
TPRS, we also discuss general parameter estimation, including population
proportion, variance, and quantiles, beyond the population mean of
the study variable.}

\textcolor{black}{We focus on rejective sampling and regression adjustment
for using general auxiliary variables in two-phase sampling, but other
design and analysis strategies are also viable. Cube methods \citep{deville2004efficient}
offer a design strategy extending to two-phase settings, ensuring
first-order inclusion probabilities for an unbiased sample although
not controlling auxiliary variable total discrepancies between phases.
Best-Choice Rerandomization \citep{wang2023asymptotic} offers a design
strategy for rejective sampling by repeatedly sampling and selecting
the sample with the best covariate balance. For other example of analysis
strategies, calibration weighting, proposed by \citet{estevao2010new}
for two-phase sampling, aligns with the two-phase regression estimator
when using generalized least squares distance. \citet{deville1992calibration}
and \citet{deville1993generalized} demonstrate that calibration estimators
are asymptotically equivalent across various distance metrics. Based
on this, we conjecture that calibration estimators in two-phase sampling,
with or without rejective sampling, share similar limiting distributions.
Also, calibration estimators complement regression adjustments by
ensuring positive or bounded weights \citep{deville2004efficient}.
Due to the complexity of deriving an analytic expression for joint
inclusion probabilities, \citet{deville2005variance} propose a general
approximation of variance estimation based on the residual technique.
We have applied similar techniques, allowing for soft calibration
instead of hard calibration, and accommodating multi-phase sampling
and general parameter estimation. }

The paper proceeds as follows. Section \ref{sec:A-review-of} provides
a review of existing design and analysis strategies of using auxiliary
variables. Sections \ref{sec:Proposed-method} and \ref{sec:General=000020pi=000020estimator}
discuss TPRS with simple random sampling and general sampling, respectively.
Section \ref{sec:Extensions} provides several extensions. Section
\ref{sec:Simulation} reports simulation results and an application
that illustrates the finite-sample performance of TPRS. We relegate
all technical details and proofs to the supplementary material.

\section{A review of design and analysis strategies of using auxiliary variables\protect\label{sec:A-review-of}}

Consider a finite population with a known size $N$. For each unit
$i$, $x_{i}$ is a $p$-dimensional auxiliary variable, and $y_{i}$
is the study variable of interest. \textcolor{black}{We focus on a
scalar $y$, but our theory extends to vector $y$ immediately.}\textcolor{blue}{{}
}The finite population quantities $\F_{N}=\{(x_{1},y_{1}),(x_{2},y_{2}),\ldots,(x_{N},y_{N})\}$
are fixed. For simplicity, we suppress the subscript $N$ on $\F$
when there is no ambiguity. The parameter of interest is the finite
population mean of the study variable $\bar{y}_{0}=N^{-1}\sum_{i=1}^{N}y_{i}$.

Two-phase sampling offers an efficient and economical method for conducting
large-scale surveys. In this section, we review the existing methods
for using auxiliary variables in two-phase sampling to enhance estimation
efficiency and identify areas needing new strategies.

\subsection{Existing design strategy: two-phase stratified sampling\protect\label{subsec:Existing-design-strategy}}

During the design stage, auxiliary variables are incorporated by selecting
a second-phase sample through stratified sampling. Strata are based
on first-phase variables, either directly from discrete variable categories
or via discretization of continuous variables. We define $x_{i}=(x_{1i},\ldots,x_{Hi})$
as the stratum indicator vector, where $x_{hi}=1$ if unit $i$ is
in stratum $h$, and $x_{hi}=0$ otherwise. Two-phase stratified sampling
proceeds as follows: 
\begin{description}
\item [{Step$\ 1.$}] From the population $\mathcal{F}$, select a first-phase
sample $\cA$ of size $n_{\I}$. For $i\in\cA,$ record $x_{i}$.
Define the sample size in each stratum $h$ as $m_{h}$ for $h=1,\ldots,H$.
The total first-phase sample size is $n_{\I}=\sum_{h=1}^{H}m_{h}$. 
\item [{Step$\ 2.$}] From each stratum $h$, randomly select $r_{h}$
units, independently across strata, as the second-phase sample $\cB$.
For $i\in\cB,$ record the study variable $y_{i}$. The total second-phase
sample size is $n_{\II}=\sum_{h=1}^{H}r_{h}$. 
\end{description}
In two-phase sampling, \textit{the double-expansion estimator} ($\pi^{*}$
estimator; \citealp{sarndal2003model}) and \textit{the reweighted
expansion estimator} (REE; \citealp{oh1983weighting,kott1997can})
are the canonical estimators for the population mean. The $\pi^{*}$
estimator,

\[
\hat{y}_{\pi^{*}}=\frac{1}{N}\sum_{i\in\cB}\frac{y_{i}}{\pi_{\II i}^{*}},
\]
where $\pi_{\II i}^{*}=\pi_{\I i}\pi_{\II i\mid\cA}$ with $\pi_{\I i}=P(i\in\cA)$
and $\pi_{\II i\mid\cA}=P(i\in\cB\mid i\in\cA)$, adjusts individual
observations by the product of their inclusion probabilities in both
phases. We can also replace $N$ in $\hat{y}_{\pi^{*}}$ by $\sum_{i\in\cB}(\pi_{\II i}^{*})^{-1}$,
emulating the Hajek estimator. However, the combined probability $\pi_{\II i}^{*}=\pi_{\I i}\pi_{\II i\mid\cA}$
is generally not the unconditional probability of a unit being in
the phase-$\II$ sample $\pr(i\in\cB)=\sum_{\cA}\pi_{\I i}\pi_{\II i\mid\cA}\pr(\cA)$,
the average probability across all possible first-phase samples, unless
$\pi_{\II i\mid\cA}$ is invariant of the first-phase sample \citep{fuller2009sampling,beaumont2016note}.
The REE,

\[
\widehat{y}_{\textsc{ree}}=\frac{1}{N}\sum_{h=1}^{H}\left(\sum_{i\in\cA}\frac{x_{hi}}{\pi_{\I i}}\right)\frac{\sum_{i\in\cB}(\pi_{\II i}^{*})^{-1}x_{hi}y_{i}}{\sum_{i\in\cB}(\pi_{\II i}^{*})^{-1}x_{hi}},
\]
recalculates the mean estimate for each stratum by a modified ratio
estimator, where the stratum-specific mean of $x$ is approximated
based on phase-$\I$ data, while the coefficient is derived from phase-$\II$
data. These estimators are often more efficient than the sample mean
estimator under simple random sampling with the same size $n_{\II}$. 

\subsection{Existing analysis strategy: regression adjustment}

\textcolor{black}{Using auxiliary variables in the analysis phase
of two-phase sampling can improve estimation efficiency. These variables
can be discrete, continuous, or a combination. The REE is one strategy
for utilizing discrete auxiliary variables. More broadly, regression
adjustment allows for utilizing general auxiliary variables, not limited
to discrete types. }

In simple random sampling across both phases, we observe variable
$x$ in phase I and variables $(x,y)$ in phase II. Then we can calculate
the mean estimates $\bar{x}_{\I}=n_{\I}^{-1}\sum_{i\in\cA}x_{i}$
from phase I and $(\bar{x}_{\II},\bar{y}_{\II})=n_{\II}^{-1}\sum_{i\in\cB}(x_{i},y_{i})$
from phase II. The two-phase regression estimator is 
\[
\bar{y}_{\reg}=\bar{y}_{\II}-(\bar{x}_{\II}-\bar{x}_{\I})^{\T}\hat{\beta}_{\II},
\]
where 
\begin{equation}
\hat{\beta}_{\II}=\left\{ \sum_{i\in\cB}\left(x_{i}-\bar{x}_{\II}\right)\left(x_{i}-\bar{x}_{\II}\right)^{\T}\right\} ^{-1}\sum_{i\in\cB}\left(x_{i}-\bar{x}_{\II}\right)\left(y_{i}-\bar{y}_{\II}\right)\label{eq:hat-beta-II}
\end{equation}
is the regression coefficient based on the phase-II sample.

The regression estimator $\bar{y}_{\reg}$ can also be written as
a weighted average of the second-phase outcomes, where the corresponding
weights are called the generalized regression estimation weights.
The regression estimator provides improved efficiency over $\bar{y}_{\II}$
in large samples if $x_{i}$ is predictive of $y_{i}$. However, issues
like large numbers of $x$-variables or imbalance between $\bar{x}_{\II}$
and $\bar{x}_{\I}$ can lead to extreme or negative weights and affect
the performance of $\bar{y}_{\reg}$ in finite samples. Researchers
have explored various methods to mitigate the issue of negative weights
associated with regression estimators, such as balanced sampling or
rejective sampling, but these approaches have mainly been studied
in single-phase sampling (\citealp{yates1946review}; \citealp{valliant2000finite};
\citealp{deville2004efficient}; \citealp{tille2005optimal}; \citealp{fuller2009some}).

We have reviewed existing design and analysis strategies in multi-phase
sampling for better efficiency, but currently it lacks strategies
for integrating both continuous and discrete variables in the design
stage. We will explore rejective sampling in multi-phase sampling
to fill the gap.

\section{TPRS: two-phase rejective sampling\protect\label{sec:Proposed-method}}

\subsection{Setup and notation}

Throughout this paper, we use $u$ and $v$ to be generic notation
for variables, which can be components of either $x$ or $y$. Define
$E(\cdot\mid\F)$, $\cov(\cdot\mid\F)$, and $\var(\cdot\mid\F)$
as expectation, covariance, and variance under the sampling design.
The finite population mean for $u_{i}$ and covariance for $u_{i}$
and $v_{i}$ are 
\begin{equation}
\bar{u}_{0}=\frac{1}{N}\sum_{i=1}^{N}u_{i},\ V_{uv,0}=\frac{1}{N-1}\sum_{i=1}^{N}(u_{i}-\bar{u}_{0})(v_{i}-\bar{v}_{0})^{\T}.\label{eq:V_uvN}
\end{equation}
The phase-I sample mean for $u_{i}$ and covariance for $u_{i}$ and
$v_{i}$ are

\begin{equation}
\bar{u}_{\I}=\frac{1}{n_{\I}}\sum_{i\in\cA}u_{i},\quad V_{uv,\I}=\frac{1}{n_{\I}-1}\sum_{i\in\cA}(u_{i}-\bar{u}_{\I})(v_{i}-\bar{v}_{\I})^{\T}.\label{eq:V_uv1}
\end{equation}
We assume $V_{xx,0}$ and $V_{xx,\I}$ are positive definite. \textcolor{black}{} 

We introduce general notation to be used throughout the paper. For
a generic vector $v$, let $v^{\otimes2}$ denote $vv^{\T}$. Define
$O_{P}(1)$ as the random variable bounded in probability, and $o_{P}(1)$
as the random variable that converges to zero in probability as $N$
increases. Let $z_{\alpha}$ represent the $100\alpha$th quantile
of the standard normal distribution.

\subsection{TPRS with simple random sampling}

Rejective sampling aims for balanced phase-II sample selection by
comparing the mean differences between phase-II and phase-I samples.
We define TPRS with simple random sampling as follows.

\begin{definition}[TPRS with simple random sampling]\label{def:TPRS}TPRS
with simple random sampling consists of two steps: 
\begin{description}
\item [{Step$\ 1.$}] Select a phase-$\I$ sample $\cA$ of size $n_{\I}$
by simple random sampling. For $i\in\cA,$ record $x_{i}$. 
\item [{Step$\ 2.$}] Select a phase-$\II$ sample $\cB$ of size $n_{\II}\leq n_{\I}$
by simple random sampling from phase-$\I$ sample $\cA$. Accept the
phase-$\II$ sample if 
\[
Q_{\I}=(\bar{x}_{\II}-\bar{x}_{\I})^{\T}\left\{ (n_{\II}^{-1}-n_{\I}^{-1})V_{xx,\I}\right\} ^{-1}(\bar{x}_{\II}-\bar{x}_{\I})<\gamma^{2},
\]
where $\gamma^{2}>0$ is a pre-specified constant, and $(n_{\II}^{-1}-n_{\I}^{-1})V_{xx,\I}$
is the phase-$\II$ design variance of $\bar{x}_{\II}-\bar{x}_{\I}$
given $\cA$. For $i\in\cB,$ record $y_{i}$. 
\end{description}
\end{definition}

In Definition \ref{def:TPRS}, if $Q_{\I}$ is below a specified threshold
$\gamma^{2}$, the phase-II sample is considered balanced and accepted
for further analysis. Based on the TPRS, the sample mean estimator
for the population mean $\bar{y}_{0}$ is 
\[
\bar{y}_{\II}=n_{\II}^{-1}\sum_{i\in\cB}y_{i}.
\]

\subsection{Large-sample design properties}

We investigate the asymptotic design-based property of $\bar{y}_{\II}$
under TPRS in Definition \ref{def:TPRS}. For asymptotic inference,
we adopt the framework of \citet{isaki1982survey}, which establishes
the asymptotic properties of estimators within a fixed sequence of
populations and corresponding random samples. This involves a series
of nested finite populations $(\mathcal{F}_{N_{1}}\subset\mathcal{F}_{N_{2}}\subset\mathcal{F}_{N_{3}}\subset\cdots)$
and sequences of samples with increasing sample sizes $(\cA_{n_{\I,1}}\subset\cA_{n_{\I,2}}\subset\cA_{n_{\I,3}}\subset\cdots$
and $\cB_{n_{\II,1}}\subset\cB_{n_{\II,2}}\subset\cB_{n_{\II,3}}\subset\cdots$).
For simplicity, we will not explicitly mention the dependence of $N_{t}$,
$n_{\I,t}$, and $n_{\II,t}$ on $t$, and refer to the asymptotic
regime as the scenario where index $t$ or $N$ goes to infinity. 

The linear projection of $y_{i}$ onto $x_{i}$ in the finite population
$\mathcal{F}$ is $\bar{y}_{0}+(x_{i}-\bar{x}_{0})^{\T}\beta_{0}$,
where 
\[
\beta_{0}=\arg\min_{\beta}\sum_{i=1}^{N}\left\{ y_{i}-\bar{y}_{0}-\left(x_{i}-\bar{x}_{0}\right)^{\T}\beta\right\} ^{2},
\]
which equals 
\begin{equation}
\beta_{0}=\left\{ \sum_{i=1}^{N}(x_{i}-\bar{x}_{0})^{\otimes2}\right\} ^{-1}\sum_{i=1}^{N}(x_{i}-\bar{x}_{0})(y_{i}-\bar{y}_{0})=V_{xx,0}^{-1}V_{xy,0}.\label{eq:betaN}
\end{equation}

\begin{assumption}\label{asmp:sampling}

Assume $\F$ contains IID samples from a superpopulation of $(x,y)$
with two conditions: (i) The sequence of $\F$ has finite $(4+\delta)$
moments for some $\delta>0$, implying $E\left(|y|^{4+\delta}\right)<\infty$
and $E\left(||x||^{4+\delta}\right)<\infty$ with respect to the superpopulation
model; (ii) $\lim_{N\rightarrow\infty}n_{\I}/N=f_{\I,0}$ and $\lim_{N\rightarrow\infty}n_{\II}/n_{\I}=f_{\II,\I}$
for some $0\leq f_{\I,0}\leq1$ and $0\leq f_{\II,\I}\leq1$. 

\end{assumption}Assumption \ref{asmp:sampling}(i) sets moment conditions
for the superpopulation, which aids in applying central limit theorems.
\citet{chen2007asymptotic} studied sufficient moment conditions on
the finite population that ensure the asymptotic normality of estimators
in two-phase sampling. Assumption \ref{asmp:sampling}(i) implies
that for any components $u$ and $v$ of $x$ and $y$, 
\begin{equation}
\lim_{N\rightarrow\infty}V_{uv,\I}=\lim_{N\rightarrow\infty}V_{uv,0}=V_{uv}\ \as,\label{eq:sigma}
\end{equation}
where $V_{uv}$ is a constant vector or matrix. Assumption \ref{asmp:sampling}(ii)
defines the sampling fractions $f_{\I,0}$ and $f_{\II,\I}$ for the
phase-I and phase-II samples, respectively. Sampling from a finite
population or phase-I sample without replacement can introduce dependency
among the individual samples. The sampling fractions serve to adjust
for such dependency when calculating the asymptotic design variances
and covariances.

Define the adjusted outcome as $e_{i}=y_{i}-x_{i}^{\T}\beta_{0}$.
The error in the sample mean estimator $\bar{y}_{\II}$ of the population
mean $\bar{y}_{0}$ decomposes into three parts: 
\begin{eqnarray}
n_{\II}^{1/2}(\bar{y}_{\II}-\bar{y}_{0}) & = & n_{\II}^{1/2}(\bar{x}_{\II}-\bar{x}_{\I})^{\T}\beta_{0}+n_{\II}^{1/2}\left\{ \bar{y}_{\II}-\bar{y}_{\I}-(\bar{x}_{\II}-\bar{x}_{\I})^{\T}\beta_{0}\right\} +n_{\II}^{1/2}(\bar{y}_{\I}-\bar{y}_{0})\nonumber \\
 & \equiv & T_{1}+T_{2}+T_{3}.\label{eq:define=000020Ts}
\end{eqnarray}
In (\ref{eq:define=000020Ts}), $T_{1}=n_{\II}^{1/2}(\bar{x}_{\II}-\bar{x}_{\I})^{\T}\beta_{0}$
and $T_{2}=n_{\II}^{1/2}(\bar{e}_{\II}-\bar{e}_{\I})$ represent the
errors of $x$ and $e$ in the phase-II sample conditional on the
phase-I sample, respectively, and $T_{3}=n_{\II}^{1/2}(\bar{y}_{\I}-\bar{y}_{0})$
represents the error of $y$ in the phase-I sample. The limiting distribution
of $(T_{1},T_{2},T_{3})$ is given in the following lemma.

\begin{lemma}\label{lemma1_srs}Suppose Assumption \ref{asmp:sampling}
holds. Without the rejection step in TPRS in Definition \ref{def:TPRS},
$(T_{1},T_{2},T_{3})$ in the decomposition (\ref{eq:define=000020Ts})
has the following limiting distribution: 
\[
\left.\left(\begin{array}{c}
T_{1}\\
T_{2}\\
T_{3}
\end{array}\right)\right\vert \F\rightarrow\N\left\{ \left(\begin{array}{c}
0\\
0\\
0
\end{array}\right),\left(\begin{array}{ccc}
\left(1-f_{\II,\I}\right)V_{yx}V_{xx}^{-1}V_{xy}, & 0 & 0\\
0 & \left(1-f_{\II,\I}\right)V_{ee} & 0\\
0 & 0 & f_{\II,\I}\left(1-f_{\I,0}\right)V_{yy}
\end{array}\right)\right\} ,
\]
$\as$ for all sequences of finite populations, where $V_{uv}$ is
defined in (\ref{eq:sigma}), and $f_{\II,\I}$ and $f_{\I,0}$ are
defined in Assumption \ref{asmp:sampling}.

\end{lemma}

With the rejection step in TPRS, the distribution of the error $n_{\II}^{1/2}(\bar{y}_{\II}-\bar{y}_{0})$
equals \textcolor{black}{the conditional distribution of }$n_{\II}^{1/2}(\bar{y}_{\II}-\bar{y}_{0})\mid(Q_{\I}<\gamma^{2})$
without rejective sampling. We use the normalized distance $D_{\I}=\{(n_{\II}^{-1}-n_{\I}^{-1})V_{xx,\I}\}^{-1/2}(\bar{x}_{\II}-\bar{x}_{\I})$
for the phase-II sample to represent the acceptance criteria $Q_{\I}<\gamma^{2}$
by $D_{\I}^{\T}D_{\I}<\gamma^{2}$. By Lemma \ref{lemma1_srs}, $D_{\I}\rightarrow\N(0,I_{p})$
and thus $D_{\I}^{\T}D_{\I}\rightarrow\chi_{p}^{2}$ a.s.. We define
the superpopulation squared correlation between $x$ and $y$ as $R^{2}=\{\corr(x,y)\}^{2}=V_{yx}V_{xx}^{-1}V_{xy}/V_{yy}$.
We show the limiting distribution of $n_{\II}^{1/2}(\bar{y}_{\II}-\bar{y}_{0})$
comprises three independent random components.

\begin{theorem}\label{Thm1}Suppose Assumption \ref{asmp:sampling}
holds. Under TPRS in Definition \ref{def:TPRS}, $\bar{y}_{\II}$
follows the limiting distribution: 
\begin{multline}
n_{\II}^{1/2}(\bar{y}_{\II}-\bar{y}_{0})\mid\left(Q_{\I}<\gamma^{2}\right)\rightarrow\left\{ (1-f_{\II,\I})V_{yy}R^{2}\right\} ^{1/2}L_{p,\gamma^{2}}\\
+\left\{ (1-f_{\II,\I})V_{yy}(1-R^{2})\right\} ^{1/2}Z_{1}+\left\{ f_{\II,\I}\left(1-f_{\I,0}\right)V_{yy}\right\} ^{1/2}Z_{2},\label{eq:rej-normal}
\end{multline}
where 
\begin{equation}
L_{p,\gamma^{2}}\sim\chi_{p,\gamma^{2}}\mathcal{S}\Gamma_{p}^{1/2},\label{eq:Lp}
\end{equation}
with $\chi_{p,\gamma^{2}}\sim\chi_{p}\mid\left(\chi_{p}^{2}\leq\gamma^{2}\right)$,
$\mathcal{S}$ follows a uniform distribution on $\{-1,1\}$, $\Gamma_{p}\sim$
$\textup{Beta}\{1/2,(p-1)/2\}$, and $\mathcal{S}\Gamma_{p}^{1/2}$
is the first coordinate of the uniform random vector over the $(p-1)$-dimensional
unit sphere, $Z_{1}$ and $Z_{2}$ are standard normal variables,
and $(L_{p,\gamma^{2}},Z_{1},Z_{2})$ are jointly independent.

\end{theorem}

The random variable $L_{p,\gamma^{2}}$, first introduced in \citet{li2016asymptotic}
for rerandomization in causal inference, is also relevant here. Our
context is more complex due to the uncertainty in phase-I estimators.

Denote $v_{p,\gamma^{2}}=\var(L_{p,\gamma^{2}})$, which equals $v_{p,\gamma^{2}}=\pr(\chi_{p+2}^{2}\leq\gamma^{2})/\pr(\chi_{p}^{2}\leq\gamma^{2})$
by (\ref{eq:Lp}) and is less than or equal to $1$ \citep{morgan2012rerandomization,li2016asymptotic}.

\begin{corollary}\label{Corollary1}Under Assumption \ref{asmp:sampling}
and TPRS in Definition \ref{def:TPRS}, the asymptotic design variance
of $n_{\II}^{1/2}(\bar{y}_{\II}-\bar{y}_{0})$ is 
\[
\left[(1-f_{\II,\I})\left\{ 1-\left(1-v_{p,\gamma^{2}}\right)R^{2}\right\} +f_{\II,\I}(1-f_{\I,0})\right]V_{yy}.
\]
The percentage reduction in asymptotic design variance compared with
the standard two-phase simple random sampling is 
\begin{equation}
\frac{1-f_{\II,\I}}{1-f_{\II,\I}f_{\I,0}}\left(1-v_{p,\gamma^{2}}\right)R^{2},\label{eq:pct=000020varred}
\end{equation}
provided $f_{\II,\I}f_{\I,0}\neq1$. If $f_{\II,\I}f_{\I,0}=1,$ indicating
a census situation, this reduction is zero. \textcolor{black}{Moreover,
$\bar{y}_{\II}$ exhibits a narrower quantile range under TPRS in
Definition \ref{def:TPRS} than under the standard two-phase simple
random sampling.}

\end{corollary}

\textcolor{black}{Without rejective sampling, $\left\{ (1-f_{\II,\I})V_{yy}R^{2}\right\} ^{1/2}L_{p,\gamma^{2}}$
in the distribution of (\ref{eq:rej-normal}) with $\gamma^{2}=\infty$
equals $\left\{ (1-f_{\II,\I})V_{yy}R^{2}\right\} ^{1/2}Z_{0}$, where
$Z_{0}$, $Z_{1}$, and $Z_{2}$ are independent standard normal variables.
When conditioned on $Q_{\I}<\gamma^{2}$, the distribution of $L_{p,\gamma^{2}}$
is more concentrated around zero than the distribution of $Z_{0}$.
This leads to a reduction in variance and quantile range for $\bar{y}_{\II}$
following rejective sampling.}

From Theorem \ref{Thm1} and Corollary \ref{Corollary1}, we discuss
the trade-off between variance and utility of the sample when selecting
$\gamma^{2}$. A lower $\gamma^{2}$ reduces the asymptotic variance
of $\bar{y}_{\II}$, and in particular, $\{(1-f_{\II,\I})V_{yy}R^{2}\}^{1/2}L_{p,\gamma^{2}}$
can be eliminated from the asymptotic distribution (\ref{eq:rej-normal})
if $\gamma$ approaches $0$ \citep{wang2022rerandomization}. In
the limit with $\gamma\rightarrow0$, the asymptotic distribution
of $\bar{y}_{\II}$ under TPRS is approximately the same as the sum
of two normal random variables $\{(1-f_{\II,\I})V_{yy}(1-R^{2})\}^{1/2}Z_{1}+\{f_{\II,\I}(1-f_{\I,0})V_{yy}\}^{1/2}Z_{2}$.
However, it also lowers the acceptance rate for the phase-II sample
and potentially limits randomness and observations in the tail of
the $x$ distribution, as discussed by\textcolor{black}{{} \citet{Legg2010}.
This reduces the utility of the sample for unplanned domain analysis,
particularly those concerning aspects associated with the tail of
the distribution of $x$. }In practice, if the primary focus is the
population parameter associated with\textcolor{black}{{} $y_{i}$},
we recommend setting\textcolor{black}{{} $\gamma^{2}$ }to the $0.001$
quantile of the \textcolor{black}{$\chi_{p}^{2}$ }distribution. This
results in an approximate acceptance rate of $0.001$, as in a related
experimental design context \textcolor{black}{\citep{li2016asymptotic}.}

Moreover, when $R^{2}=0$, indicating no correlation between auxiliary
and study variables, the rejective procedure does not affect the limiting
distribution for $\bar{y}_{\II}$. Conversely, with $R^{2}=1$, the
variance reduction for $\bar{y}_{\II}$ is maximized due to the strong
correlation between auxiliary and study variables.

\subsection{Regression estimator}

Design and analysis strategies can be integrated: \citet{fuller2009some}
combined rejective sampling and regression adjustment in surveys,
and \citet{li2020rerandomization} combined rerandomization and regression
adjustment in experiments. Under TPRS, a regression estimator is 
\begin{equation}
\bar{y}_{\II,\reg}=\bar{y}_{\II}-\left(\bar{x}_{\II}-\bar{x}_{\I}\right)^{\T}\hat{\beta}_{\II},\label{eq:2preg}
\end{equation}
with $\hat{\beta}_{\II}$ specified earlier in (\ref{eq:hat-beta-II}).
To derive the asymptotic distribution of $\bar{y}_{\II,\reg}$ in
TPRS, we decompose it as follows 
\begin{eqnarray}
n_{\II}^{1/2}\left(\bar{y}_{\II,\reg}-\bar{y}_{0}\right) & = & n_{\II}^{1/2}\left(\bar{y}_{\II}-\bar{y}_{\I}\right)-n_{\II}^{1/2}(\bar{x}_{\II}-\bar{x}_{\I})^{\T}\beta_{0}\nonumber \\
 &  & -n_{\II}^{1/2}(\bar{x}_{\II}-\bar{x}_{\I})^{\T}(\hat{\beta}_{\II}-\beta_{0})+n_{\II}^{1/2}\left(\bar{y}_{\I}-\bar{y}_{0}\right)\nonumber \\
 & = & n_{\II}^{1/2}\left(\bar{e}_{\II}-\bar{e}_{\I}\right)-n_{\II}^{1/2}(\bar{x}_{\II}-\bar{x}_{\I})^{\T}(\hat{\beta}_{\II}-\beta_{0})+n_{\II}^{1/2}\left(\bar{y}_{\I}-\bar{y}_{0}\right).\label{eq:A4}
\end{eqnarray}
Here, $n_{\II}^{1/2}(\bar{e}_{\II}-\bar{e}_{\I})$ represents the
error of $e$ in the phase-II sample conditional on the phase-I sample,
$n_{\II}^{1/2}(\bar{x}_{\II}-\bar{x}_{\I})^{\T}(\hat{\beta}_{\II}-\beta_{0})$
is of order $o_{P}(1)$, and $n_{\II}^{1/2}(\bar{y}_{\I}-\bar{y}_{0})$
represents the error of $y$ in the phase-I sample. The limiting distribution
of $\bar{y}_{\II,\reg}$ is given in the following theorem.

\begin{theorem}\label{thm:2preg-rej}Suppose Assumption \ref{asmp:sampling}
holds. Under TPRS in Definition \ref{def:TPRS}, the regression estimator
$\bar{y}_{\II,\reg}$ in (\ref{eq:2preg}) has the following limiting
distribution:

\begin{equation}
n_{\II}^{1/2}(\bar{y}_{\II,\reg}-\bar{y}_{0})\mid\F\rightarrow\N\left\{ 0,f_{\II,\I}(1-f_{\I,0})V_{yy}+(1-f_{\II,\I})V_{ee}\right\} ,\label{eq:asympReg}
\end{equation}
$\as$, where $V_{uv}$ is defined in (\ref{eq:sigma}), $e=y-x^{\T}\beta_{0}$,
and $f_{\II,\I}$ and $f_{\I,0}$ are defined in Assumption \ref{asmp:sampling}.

\end{theorem}

\citet{fuller2009some} provided the consistency and asymptotic variance
of the regression estimator under single phase rejective sampling,
but did not include the asymptotic distribution results. The asymptotic
distribution result in Theorem \ref{thm:2preg-rej} complements \citet{fuller2009some}'s
work, specifically by considering a census in phase $\I$.

The limiting distribution of $n_{\II}^{1/2}(\bar{y}_{\II,\reg}-\bar{y}_{0})\mid\F$
in (\ref{eq:asympReg}) under TPRS does not depend on $\gamma^{2}$,
and as $\gamma^{2}$ increases, it becomes equivalent to non-rejective
sampling. This demonstrates the equivalence of the limiting distributions
of $\bar{y}_{\II,\reg}$ with and without rejective sampling, aligning
with \citet{fuller2009some}, who found similar performance of regression
estimators in single-phase samples regardless of rejective sampling.

Theorems \ref{Thm1} and \ref{thm:2preg-rej} highlight the role of
rejective sampling in estimation. When $\gamma^{2}\approx0$, the
asymptotic design variance of the simple mean estimator $\bar{y}_{\II}$
is close to that of the two-phase regression estimator $\bar{y}_{\II,\reg}$.
That is, the simple mean estimator under rejective sampling performs
similarly to the regression estimator. 

\textcolor{black}{If the two-phase regression estimator includes additional
covariates (regressors) beyond the design covariates $x$, the results
in Theorem \ref{thm:2preg-rej} still apply by replacing $x$ with
the regressors, whether rejective sampling is used or not. }Despite
similar limiting properties with and without rejective sampling, \citet{fuller2009some}
recommended rejective sampling for its practical benefits. We can
express the regression estimator as a weighted average of the $y_{i}$'s
in the phase-II sample $\bar{y}_{\II,\reg}=\sum_{i\in\cB}\omega_{i}y_{i}$,
where 
\[
\omega_{i}=1+\left(\bar{x}_{\I}-\bar{x}_{\II}\right)^{\T}\left\{ \sum_{i\in\cB}\left(x_{i}-\bar{x}_{\II}\right)^{\otimes2}\right\} ^{-1}\left(x_{i}-\bar{x}_{\II}\right)
\]
are the weights. Without rejective sampling, $\omega_{i}$ may be
negative due to influential values of $x_{i}$, which affects the
robustness of the regression estimator. Rejective sampling reduces
this chance, as demonstrated by simulation from \citet{Legg2010}
under single-phase sampling and further confirmed by our simulation
under TPRS. \textcolor{black}{However, rejective sampling does not
completely eliminate the occurrence of negative weights. In these
cases, one might consider alternative estimator, such as calibration
weighting, as discussed in Section \ref{sec:Introduction}. }

Moreover, rejective sampling improves the covariate balances in phase-II
sample, evidenced by a smaller asymptotic design variance of $n_{\II}^{1/2}(\bar{x}_{\II}-\bar{x}_{0})$
compared with designs without rejective sampling. Specially, under
TPRS, the limit of the design covariance $\acov\{n_{\II}^{1/2}(\bar{x}_{\II}-\bar{x}_{0})\mid Q_{\I}<\gamma^{2}\}=\{(1-f_{\II,\I})v_{p,\gamma^{2}}+(f_{\II,\I}-f_{\II,0})\}V_{xx}$
is always no larger than that without rejective sampling $\acov\{n_{\II}^{1/2}(\bar{x}_{\II}-\bar{x}_{0})\mid\F\}=(1-f_{\II,0})V_{xx}$.

\subsection{Inference: variance estimators and confidence intervals\protect\label{subsec:ve-srs}}

To infer the population mean $\bar{y}_{0}$ based on Theorems \ref{Thm1}
and \ref{thm:2preg-rej}, we first estimate the asymptotic design
variances and covariances. Let the estimator for $V_{uv}$ be

\[
\hat{V}_{uv}=\frac{1}{n_{\II}-1}\sum_{i\in\cB}(u_{i}-\bar{u}_{\II})(v_{i}-\bar{v}_{\II})^{\T}.
\]

\begin{proposition}\label{prop:ve}Under Assumption \ref{asmp:sampling},
$\hat{V}_{uv}$ is a consistent estimator of $V_{uv}$ under both
two-phase simple random sampling and TPRS in Definition \ref{def:TPRS}.

\end{proposition}

A consistent estimator for $R^{2}$ is $\hat{R}^{2}=(\hat{V}_{yx}\hat{V}_{xx}^{-1}\hat{V}_{xy})/\hat{V}_{yy}$.
The variance estimator for $\bar{y}_{\II}$ under TPRS is 
\[
\frac{1}{n_{\II}}\left[\left(1-\frac{n_{\II}}{n_{\I}}\right)\left\{ 1-\left(1-v_{p,\gamma^{2}}\right)\hat{R}^{2}\right\} +\frac{n_{\II}}{n_{\I}}\left(1-\frac{n_{\I}}{N}\right)\right]\hat{V}_{yy}.
\]
Define $\hat{e}_{i}=y_{i}-x_{i}^{\T}\hat{\beta}_{\II}$ for phase-II
samples. Then, we can estimate $V_{ee}$ by 
\[
\hat{V}_{ee}=\frac{1}{n_{\II}-p-1}\sum_{i\in\cB}(\hat{e}_{i}-\bar{e}_{\II})^{2},
\]
where $n_{\II}-p-1$ adjusts for degrees of freedom due to estimating
$\beta_{0}$. Decompose $V_{yy}$ into $\beta_{0}^{\T}V_{xx}\beta_{0}+V_{ee}$
and estimate it by $\hat{V}_{yy}=\hat{\beta}_{\II}^{\T}\hat{V}_{xx}\hat{\beta}_{\II}+\hat{V}_{ee}$.
A consistent variance estimator for $\bar{y}_{\II,\reg}$ is 
\begin{equation}
\hat{V}_{\reg}=\frac{1}{n_{\II}}\left\{ \frac{n_{\II}}{n_{\I}}\left(1-\frac{n_{\I}}{N}\right)\hat{V}_{yy}+\left(1-\frac{n_{\II}}{n_{\I}}\right)\hat{V}_{ee}\right\} .\label{eq:ve=000020for=000020reg}
\end{equation}

We can construct the asymptotic $(1-\alpha)$ confidence interval
of $\bar{y}_{0}$ based on $\bar{y}_{\II}$ as 
\[
\left(\bar{y}_{\II}-n_{\II}^{-1/2}\nu_{1-\alpha/2}(\hat{R}^{2})\hat{V}_{yy}^{1/2},\ \bar{y}_{\II}-n_{\II}^{-1/2}\nu_{\alpha/2}(\hat{R}^{2})\hat{V}_{yy}^{1/2}\right),
\]
where $\nu_{\alpha}(R^{2})$ as the $100\alpha$th quantile of the
distribution of 
\[
\left(1-\frac{n_{\II}}{n_{\I}}\right)^{1/2}\left\{ RL_{p,\gamma^{2}}+\left(1-R^{2}\right)^{1/2}Z_{2}\right\} +\left\{ \frac{n_{\II}}{n_{\I}}\left(1-\frac{n_{\I}}{N}\right)\right\} ^{1/2}Z_{3},
\]
and the counterpart based on $\bar{y}_{\II,\reg}$ as 
\[
\left(\bar{y}_{\II,\reg}-n_{\II}^{-1/2}\hat{V}_{\reg}^{1/2}z_{1-\alpha/2},\ \bar{y}_{\II}-n_{\II}^{-1/2}\hat{V}_{\reg}^{1/2}z_{\alpha/2}\right).
\]

\section{TPRS with general sampling\protect\label{sec:General=000020pi=000020estimator} }

\subsection{Notation}

\textcolor{black}{Two-phase sampling is commonly used in national
health surveys, such as the US National Health and Nutrition Examination
Survey \citep{NHANES2024} and the US National Health Interview Survey
\citep{NHIS2024}. Complex sampling designs are often considered for
both first- and second-phase samples, such as Poisson sampling designs,
stratified multistage cluster sample designs and sampling with probability
proportional to measures of size.}

We now consider two-phase sampling with general phase-$\I$ and $\II$
designs. Let $\pi_{\I i}$ be the probability of including unit $i$
in the phase-$\I$ sample $\cA$, and let $\pi_{\II i\mid\cA}$ be
the conditional probability of including unit $i$ in the phase-$\II$
sample $\cB$ given that unit $i$ is in the phase-$\I$ sample. Let
$n_{\I}$ and $n_{\II}$ be the sample sizes of the phase-I sample
and the phase-II sample, respectively. 

Define the finite population mean as $\bar{u}_{0}=N^{-1}\sum_{i=1}^{N}u_{i}$.
With slight abuse of the notation, define the phase-$\I$ estimator
and the phase-II estimator as 
\begin{equation}
\bar{u}_{\I}=\frac{1}{\sum_{i\in\cA}\pi_{\I i}^{-1}}\sum_{i\in\cA}\frac{u_{i}}{\pi_{\I i}},\label{eq:pIpi}
\end{equation}
and 
\begin{equation}
\bar{u}_{\II}=\frac{1}{\sum_{i\in\cB}(\pi_{\II i}^{*})^{-1}}\sum_{i\in\cB}\frac{u_{i}}{\pi_{\II i}^{*}},\label{eq:pIIpi}
\end{equation}
respectively. The phase-I estimator is a Hájek estimator for $\bar{u}_{0}$,
while the phase-II estimator, known as the double expansion estimator
\citep{kott1997can} or a $\pi^{*}$ estimator \citep{sarndal2003model},
is generally not a Hájek estimator for $\bar{u}_{0}$ because $\pi_{\II i}^{*}=\pi_{\I i}\pi_{\II i\mid\cA}$
is not the probability of $i$ being selected for phase II in general
as we discussed in Section \ref{subsec:Existing-design-strategy}.

To calculate the design variances for (\ref{eq:pIpi}) and (\ref{eq:pIIpi}),
we require positive second-order inclusion probabilities. The probabilities,
$\pi_{\I ij}=P(i,j\in\cA\mid\F)$ and $\pi_{\II ij\mid\cA}=P(i,j\in\cB\mid i,j\in\cA)$,
determine the probability or the conditional probability of pairs
of units being included in phase-I and phase-II samples, respectively.

Under suitable regularity conditions on sampling (with details in
Section \ref{subsec:Two-phase-rejective}), the sums $\sum_{i\in\cA}\pi_{\I i}^{-1}$
and $\sum_{i\in\cB}(\pi_{\II i}^{*})^{-1}$ are design consistent
for the population size $N$. Using Taylor expansion and ignoring
small order terms, the design covariance of $\bar{u}_{\I}$ and $\bar{v}_{\I}$
is

\begin{equation}
V_{uv,0}=\cov\left(\bar{u}_{\I},\bar{v}_{\I}\mid\F\right)=\frac{1}{N^{2}}\sum_{i=1}^{N}\sum_{j=1}^{N}\frac{\pi_{\I ij}-\pi_{\I i}\pi_{\I j}}{\pi_{\I i}\pi_{\I j}}(u_{i}-\bar{u}_{0})(v_{j}-\bar{v}_{0})^{\T},\label{eq:V_uvN-pi}
\end{equation}
and the conditional design covariance of $\bar{u}_{\II}$ and $\bar{v}_{\II}$
given the phase-$\I$ sample is 
\begin{equation}
V_{uv,\I}=\cov\left(\bar{u}_{\II},\bar{v}_{\II}\mid\cA,\F\right)=\frac{1}{N^{2}}\sum_{i\in\cA}\sum_{j\in\cA}\frac{\pi_{\II ij\mid\cA}-\pi_{\II i\mid\cA}\pi_{\II j\mid\cA}}{\pi_{\II i}^{*}\pi_{\II j}^{*}}(u_{i}-\bar{u}_{\I})(v_{j}-\bar{v}_{\I})^{\T}.\label{eq:V_uv1-pi}
\end{equation}
We assume that $V_{xx,0}$ and $V_{xx,\I}$ are positive definite
for all phase-I samples.

\subsection{TPRS with general sampling\protect\label{subsec:Two-phase-rejective}}

We define TPRS with general sampling designs as follows.

\begin{definition}[TPRS with general sampling]\label{def:TPRS_gen}
TPRS with general sampling consists of two steps: 
\begin{description}
\item [{Step$\ 1.$}] Select a phase-$\I$ sample $\cA$ by a general $\pi$
sampling with the inclusion probability $\pi_{\I i}$. For $i\in\cA,$
record $x_{i}$. 
\item [{Step$\ 2.$}] Treat the phase-$\I$ sample $\cA$ as the population
and select a phase-$\II$ sample $\cB$ by a general $\pi$ sampling
with the conditional inclusion probability $\pi_{\II i\mid\cA}$.
Accept the phase-$\II$ sample if 
\[
Q_{\I}=(\bar{x}_{\II}-\bar{x}_{\I})^{\T}V_{xx,\I}^{-1}(\bar{x}_{\II}-\bar{x}_{\I})<\gamma^{2},
\]
where $\gamma^{2}>0$ is a pre-specified constant, and $V_{xx,\I}$
is the design variance of $\bar{x}_{\II}-\bar{x}_{\I}$ given $\cA$,
given by (\ref{eq:V_uv1-pi}) with $u$ and $v$ being $x$. For $i\in\cB,$
record $y_{i}$. 
\end{description}
\end{definition}

For the population mean $\bar{y}_{0}$, the $\pi^{*}$ estimator is
\begin{equation}
\bar{y}_{\II}=\frac{1}{\sum_{i\in\cB}(\pi_{\II i}^{*})^{-1}}\sum_{i\in\cB}\frac{y_{i}}{\pi_{\II i}^{*}},\label{eq:ybar-II}
\end{equation}
recalling that $\pi_{\II i}^{*}=\pi_{\I i}\pi_{\II i\mid\cA}$. We
focus on the $\pi^{*}$ estimator for simplicity because the REE reviewed
in Section \ref{subsec:Existing-design-strategy} is more natural
for two-phase stratified sampling. Below we will show that integrating
the $\pi^{*}$ estimator with TPRS suffices to attain favorable design
properties.

\subsection{Large-sample design properties}

To understand the limiting properties of $\bar{y}_{\II}$, we follow
the asymptotic framework in Section \ref{sec:Proposed-method} and
specify the following regularity conditions for TPRS in Definition
\ref{def:TPRS_gen}.

\begin{assumption}\label{asmp:sampling-pi}Assumption \ref{asmp:sampling}(i)
and (ii) hold. (iii) The phase-$\I$ estimator (\ref{eq:pIpi}) satisfies
\[
\var(\bar{u}_{\I}\mid\F)^{-1/2}(\bar{u}_{\I}-\bar{u}_{0})\mid\F\rightarrow\N(0,1)\ \as,
\]
with $\var(n_{\I}^{1/2}\bar{u}_{\I}\mid\F)=O_{P}(1)$, where $u$
represents components of either $x$ or $y$.

(iv) The sequence of phase-$\I$ selection probabilities are bounded
by $K_{\I,\textup{L}}<n_{\I}^{-1}N\pi_{\I i}<K_{\I,\textup{U}}$ for
all $i$, for some positive $K_{\I,\textup{L}}>0$ and $K_{\I,\textup{U}}>0$,
and the design weighted sums of moments converge to constants, 
\[
\lim_{N\rightarrow\infty}\sum_{i\in\cA}\pi_{\I i}^{-1}(1,x_{i}^{\T},y_{i},y_{i}^{2})^{\T}(1,x_{i}^{\T},y_{i},y_{i}^{2})=M_{\I}\ \as,
\]
where $M_{\I}$ is a matrix of constants.

(v) The phase-$\II$ estimator (\ref{eq:pIIpi}) satisfies 
\[
\var(\bar{u}_{\II}\mid\cA,\F)^{-1/2}(\bar{u}_{\II}-\bar{u}_{\I})\mid\cA,\F\rightarrow\N(0,1)\ \as,
\]
and $\var(n_{\II}^{1/2}\bar{u}_{\II}\mid\cA,\F)=O_{P}(1)$;

(vi) The sequence of phase-$\II$ selection probabilities are bounded
by $K_{\II,\textup{L}}<n_{\II}^{-1}n_{\I}\pi_{\II i\mid\cA}<K_{\II,\textup{U}}$
for all $i$, for some positive $K_{\II,\textup{L}}>0$ and $K_{\II,\textup{U}}>0$,
and the design weighted sums of moments converge to constants, 
\[
\lim_{N\rightarrow\infty}\sum_{i\in\cB}\pi_{\II i\mid\cA}^{-1}(1,x_{i}^{\T},y_{i},y_{i}^{2})^{\T}(1,x_{i}^{\T},y_{i},y_{i}^{2})=M_{\II}\ \as,
\]
where $M_{\II}$ is a matrix of constants.

(vii) The design covariance between the differences in $x$ and $e$
in phases I and II is negligible: $\cov(\bar{x}_{\II}-\bar{x}_{\I},\bar{e}_{\II}-\bar{e}_{\I}\mid\cA,\F)=o_{P}(n_{\II}^{-1})$.

\end{assumption}

The conditions in Assumption \ref{asmp:sampling-pi} are standard
for sample moments and sampling designs (\citealp{fuller2009sampling},
Theorem 3.3.1). They ensure the general applicability of the phase-I
and phase-II estimators across various designs (Chapter 3, \citealp{fuller2009sampling}).
For instance, Assumption \ref{asmp:sampling-pi}(vii) holds under
two-phase simple random sampling and stratified sampling. A heuristic
explanation is provided below. Let $r_{i}=y_{i}-x_{i}^{\T}\hat{\beta}_{\I}$
be the residual based on the phase-I regression had the study variables
been measured in the phase-I sample, where\textcolor{red}{{} }$\hat{\beta}_{\I}=\{\sum_{i\in\cA}\pi_{\I i}^{-1}(x_{i}-\bar{x}_{\I})^{\otimes2}\}^{-1}\sum_{i\in\cA}\pi_{\I i}^{-1}(x_{i}-\bar{x}_{\I})(y_{i}-\bar{y}_{\I})$
is the phase-I regression coefficient. Then, we have $\cov(\bar{x}_{\II}-\bar{x}_{\I},\bar{r}_{\II}-\bar{r}_{\I}\mid\cA,\F)=0$.
Recall that $e_{i}=y_{i}-x_{i}^{\T}\beta_{0}$, which can be written
as $e_{i}=r_{i}+x_{i}^{\T}(\hat{\beta}_{\I}-\beta_{0})$. Thus, we
have 
\begin{eqnarray}
\cov(\bar{x}_{\II}-\bar{x}_{\I},\bar{e}_{\II}-\bar{e}_{\I}\mid\cA,\F) & = & \cov(\bar{x}_{\II}-\bar{x}_{\I},\bar{r}_{\II}-\bar{r}_{\I}\mid\cA,\F)\nonumber \\
 &  & +\cov\{\bar{x}_{\II}-\bar{x}_{\I},(\bar{x}_{\II}-\bar{x}_{\I})^{\T}(\hat{\beta}_{\I}-\beta_{0})\mid\cA,\F\}\nonumber \\
 & = & \var(\bar{x}_{\II}-\bar{x}_{\I}\mid\cA,\F)(\hat{\beta}_{\I}-\beta_{0}).\label{eq:A(vii)}
\end{eqnarray}
Because $\var(\bar{x}_{\II}-\bar{x}_{\I}\mid\cA,\F)=O_{P}(n_{\II}^{-1})$
$\as$, where the probability distribution in $O_{P}$ is induced
by phase-II sampling given $(\cA,\F)$, and $\hat{\beta}_{\I}-\beta_{0}=O_{P}(n_{\I}^{-1})$,
where the probability distribution in $O_{P}$ is induced by phase-I
sampling, the quantity in (\ref{eq:A(vii)}) is of order $o_{P}(n_{\II}^{-1})$. 

Following the decomposition in (\ref{eq:define=000020Ts}), we decompose
$n_{\II}^{1/2}(\bar{y}_{\II}-\bar{y}_{0})$ into three components
\begin{equation}
n_{\II}^{1/2}\left(\bar{y}_{\II}-\bar{y}_{0}\right)=T_{1}+T_{2}+T_{3},\label{eq:decomposition-pi}
\end{equation}
where $T_{1}=n_{\II}^{1/2}(\bar{x}_{\II}-\bar{x}_{\I})^{\T}\beta_{0}$,
$T_{2}=n_{\II}^{1/2}(\bar{e}_{\II}-\bar{e}_{\I})$, and $T_{3}=n_{\II}^{1/2}(\bar{y}_{\I}-\bar{y}_{0})$
with the general $\pi$ estimators for $\bar{u}_{\I}$ and $\bar{u}_{\II}$
defined in (\ref{eq:pIpi}) and (\ref{eq:pIIpi}), respectively. The
limiting distribution of $(T_{1},T_{2},T_{3})$ is given in the following
lemma.

\begin{lemma}\label{Lemma2-pi}Suppose Assumption \ref{asmp:sampling-pi}
holds. Without the rejection step in TPRS in Definition \ref{def:TPRS_gen},
the joint distribution of $(T_{1},T_{2},T_{3})$ in the decomposition
(\ref{eq:decomposition-pi}) has the following limiting distribution:

\[
\left.\left(\begin{array}{c}
T_{1}\\
T_{2}\\
T_{3}
\end{array}\right)\right\vert \F\rightarrow\N\left\{ \left(\begin{array}{c}
0\\
0\\
0
\end{array}\right),\left(\begin{array}{ccc}
V_{1}, & 0 & 0\\
0 & V_{2} & 0\\
0 & 0 & V_{3}
\end{array}\right)\right\} ,
\]
$\as$ for all sequences of finite populations, where 
\begin{eqnarray}
V_{1} & = & \lim_{N\rightarrow\infty}n_{\II}\beta_{0}^{\T}E\left(V_{xx,\I}\mid\F\right)\beta_{0},\label{eq:V1}\\
V_{2} & = & \lim_{N\rightarrow\infty}n_{\II}E\left(V_{ee,\I}\mid\F\right),\label{eq:V2}\\
V_{3} & = & \lim_{N\rightarrow\infty}n_{\II}V_{yy,0}.\label{eq:V3}
\end{eqnarray}

\end{lemma}

By construction, the distribution of $n_{\II}^{1/2}(\bar{y}_{\II}-\bar{y}_{0})$
under TPRS in Definition \ref{def:TPRS_gen} is equivalent to the
distribution of $n_{\II}^{1/2}(\bar{y}_{\II}-\bar{y}_{0})\mid(Q_{\I}<\gamma^{2})$
without rejective sampling. To study the asymptotic design property,
we define $D_{\I}=(n_{\II}V_{xx,\I})^{-1/2}n_{\II}^{1/2}(\bar{x}_{\II}-\bar{x}_{\I})$.
Then, $Q_{\I}$ is expressed as $D_{\I}^{\T}D_{\I}$, with $D_{\I}\rightarrow\N(0,I_{p})$
and thus $D_{\I}^{\T}D_{\I}\rightarrow\chi_{p}^{2}$ a.s..

\begin{theorem}\label{Thm1-1}Suppose Assumption \ref{asmp:sampling-pi}
holds. Under TPRS  in Definition \ref{def:TPRS_gen}, $\bar{y}_{\II}$
follows the limiting distribution: \textcolor{black}{
\begin{equation}
n_{\II}^{1/2}(\bar{y}_{\II}-\bar{y}_{0})\mid\left(Q_{\I}<\gamma^{2}\right)\rightarrow V_{1}^{1/2}L_{p,\gamma^{2}}+V_{2}^{1/2}Z_{1}+V_{3}^{1/2}Z_{2},\label{eq:rej-normal-1}
\end{equation}
}where $V_{1}$, $V_{2},$ and $V_{3}$ are defined in (\ref{eq:V1})--(\ref{eq:V3}),
$Z_{1}$ and $Z_{2}$ are standard normal variables, and $(L_{p,\gamma^{2}},Z_{1},Z_{2})$
are jointly independent.

\end{theorem}

The results of TPRS with general sampling are similar to those of
TPRS with simple random sampling, hence inheriting all the benefits,
including enhanced covariate balance, and reduced variance and quantile
range, as detailed in Section \ref{sec:Proposed-method}.

\subsection{Regression estimator}

Integrating the design and analysis strategies, the two-phase regression
estimator for $\bar{y}_{0}$ is 
\begin{equation}
\bar{y}_{\II,\reg}=\bar{y}_{\II}-\left(\bar{x}_{\II}-\bar{x}_{\I}\right)^{\T}\hat{\beta}_{\II},\label{eq:2preg-pi}
\end{equation}
where 
\[
\hat{\beta}_{\II}=\left\{ \sum_{i\in\cB}\frac{\left(x_{i}-\bar{x}_{\II}\right)^{\otimes2}}{\pi_{\II i}^{*}}\right\} ^{-1}\sum_{i\in\cB}\frac{(x_{i}-\bar{x}_{\II})(y_{i}-\bar{y}_{\II})}{\pi_{\II i}^{*}}
\]
is the regression coefficient based on the phase-II sample.

\begin{theorem}\label{Thm:reg-pi}Suppose Assumption \ref{asmp:sampling-pi}
holds. Under TPRS in Definition \ref{def:TPRS_gen}, the regression
estimator $\bar{y}_{\II,\reg}$ in (\ref{eq:2preg-pi}) has the following
limiting distribution:
\[
n_{\II}^{1/2}(\bar{y}_{\II,\reg}-\bar{y}_{0})\mid\F\rightarrow\N\left(0,V_{\II,\reg}\right),
\]
$\as$ for all sequences of finite populations, where 
\[
V_{\II,\reg}=\lim_{N\rightarrow\infty}n_{\II}\left\{ V_{yy,0}+E\left(V_{ee,\I}\mid\F\right)\right\} 
\]
with $V_{yy,0}$ and $V_{ee,\I}$ defined in (\ref{eq:V_uvN-pi})
and (\ref{eq:V_uv1-pi}), respectively.

\end{theorem}

Theorem \ref{Thm:reg-pi} indicates that the advantages of TPRS are
also applicable to the regression estimator in a general setup.

\subsection{Inference: variance estimators and confidence intervals\protect\label{subsec:ve-pi}}

We estimate the asymptotic design variances of $\bar{y}_{\II}$ and
$\bar{y}_{\II,\reg}$ in general TPRS. For $\bar{y}_{\II}$, the variance
is estimated by $n_{\II}^{-1}(\hat{V}_{1}v_{p,\gamma^{2}}+\hat{V}_{2}+\hat{V}_{3}),$
where 
\begin{eqnarray}
\hat{V}_{1} & = & n_{\II}\hat{\beta}_{\II}^{\T}V_{xx,\I}\hat{\beta}_{\II},\nonumber \\
\hat{V}_{2} & = & \frac{n_{\II}}{N^{2}}\sum_{i\in\cB}\sum_{j\in\cB}\frac{\pi_{\II ij\mid\cA}-\pi_{\II i\mid\cA}\pi_{\II j\mid\cA}}{\pi_{\II i}^{*}\pi_{\II j}^{*}}\frac{(\hat{e}_{i}-\hat{e}_{\II})(\hat{e}_{j}-\hat{e}_{\II})^{\T}}{\pi_{\II ij\mid\cA}},\label{eq:V-hat2}\\
\hat{V}_{3} & = & \frac{n_{\II}}{N^{2}}\sum_{i\in\cB}\sum_{j\in\cB}\frac{\pi_{\I ij}-\pi_{\I i}\pi_{\I j}}{\pi_{\I i}\pi_{\I j}}\frac{(y_{i}-\bar{y}_{\II})(y_{j}-\bar{y}_{\II})^{\T}}{\pi_{\I ij}\pi_{\II ij\mid\cA}},\label{eq:V-hat3}
\end{eqnarray}
and $\hat{e}_{i}=y_{i}-x_{i}^{\T}\hat{\beta}_{\II}$ and $\hat{e}_{\II}$
are (\ref{eq:ybar-II}) with $y_{i}$ being $\hat{e}_{i}$. For $\bar{y}_{\II,\reg}$,
the variance is estimated by $n_{\II}^{-1}(\hat{V}_{2}+\hat{V}_{3}).$

\begin{remark}\label{Remark:SYG}Regarding variance estimation in
two-phase sampling, two points are noteworthy. First, obtaining the
joint inclusion probabilities $\pi_{\I ij}$ and $\pi_{\II ij\mid\cA}$
may be difficult in practice, and approximations may be needed. For
example, \citet{haziza2005estimation} and \citet{beaumont2015clarifying}
considered a simplified variance estimator which does not require
the information on $\pi_{\II ij\mid\cA}$ and specified conditions
under which the bias of the variance estimator is negligible.

Second, while Horvitz--Thompson-type variance estimators (\ref{eq:V-hat2})
and (\ref{eq:V-hat3}) are standard, they can be unstable and negative
in unequal probability sampling. An alternative is the Sen--Yates--Grundy-type
(\citealp{sen1953estimate,yates1953selection}) variance estimators
\begin{eqnarray}
\hat{V}_{2,\mathrm{SYG}} & = & -\frac{1}{2}\frac{n_{\II}}{N^{2}}\sum_{i\in\cB}\sum_{j\in\cB}\frac{\pi_{\II ij\mid\cA}-\pi_{\II i\mid\cA}\pi_{\II j\mid\cA}}{\pi_{\II ij\mid\cA}}\left(\frac{\hat{e}_{i}-\hat{e}_{\II}}{\pi_{\II i}^{*}}-\frac{\hat{e}_{j}-\hat{e}_{\II}}{\pi_{\II j}^{*}}\right)^{\otimes2},\label{eq:V-hat2-syg}\\
\hat{V}_{3,\mathrm{SYG}} & = & -\frac{1}{2}\frac{n_{\II}}{N^{2}}\sum_{i\in\cB}\sum_{j\in\cB}\left(\pi_{\I ij}-\pi_{\I i}\pi_{\I j}\right)\left(\frac{y_{i}-\bar{y}_{\II}}{\pi_{\I i}}-\frac{y_{j}-\bar{y}_{\II}}{\pi_{\I j}}\right)^{\otimes2}.\label{eq:V-hat3-syg}
\end{eqnarray}
These estimators are asymptotically equivalent to $\hat{V}_{2}$ and
$\hat{V}_{3}$ with additional mean zero terms but are more stable
and non-negative for various sampling designs with fixed sample sizes
\citep{hidiroglou2009variance}.

\end{remark}

We can construct the asymptotic $(1-\alpha)$ confidence interval
for $\bar{y}_{0}$ based on $\bar{y}_{\II}$ as
\[
\left(\bar{y}_{\II}-n_{\II}^{-1/2}\nu_{1-\alpha/2}(\hat{V}_{1},\hat{V}_{2},\hat{V}_{3}),\ \bar{y}_{\II}-n_{\II}^{-1/2}\nu_{\alpha/2}(\hat{V}_{1},\hat{V}_{2},\hat{V}_{3})\right),
\]
where $\nu_{\alpha}(V_{1},V_{2},V_{3})$ as the $100\alpha$th quantile
of the distribution of $V_{1}^{1/2}L_{p,\gamma^{2}}+V_{2}^{1/2}Z_{1}+V_{3}^{1/2}Z_{2}$,
and the counterpart based on $\bar{y}_{\II,\reg}$ as
\[
\left(\bar{y}_{\II,\reg}-n_{\II}^{-1/2}\left(\hat{V}_{2}+\hat{V}_{3}\right)^{1/2}z_{1-\alpha/2},\ \bar{y}_{\II,\reg}-n_{\II}^{-1/2}\left(\hat{V}_{2}+\hat{V}_{3}\right)^{1/2}z_{\alpha/2}\right).
\]

\section{Extensions\protect\label{sec:Extensions}}

\subsection{Other designs}

When covariates have varying importance, applying different thresholds
for different covariates in rejective sampling can be more effective.
\citet{Legg2010} explored two approaches for this: \textcolor{black}{weighted
rejective sampling} and sequential rejective sampling, with the former
also studied by \citet{lu2023design} and the latter by \citet{morgan2015rerandomization}
and \citet{li2016asymptotic} in rerandomization in experiments. We
further develop sequential rejective sampling in two-phase sampling
in Section \ref{sec:Sequential-rejective-sampling} of the supplementary
material and demonstrate that it provides better balance control for
each specific covariate compared with weighted rejective sampling. 

Moreover, we introduce multi-phase rejective sampling and establish
asymptotic design properties of the double expansion estimator and
the regression estimator in Section \ref{sec:Multi-phase-rejective-sampling}
of the supplementary material. As a special case, if phase I is a
census, the three-phase rejective sampling reduces to TPRS with rejective
sampling in both phases. In both sequential and multi-phase rejective
sampling methods, we employ block-wise Gram--Schmidt orthogonalization
on tiers or phases of covariates. This strategy ensures a clear distinction
between various sets of covariates, categorizing them according to
their level of importance or their availability across different phases.

\subsection{General parameters}

\textcolor{black}{The current framework can be extended to deal with
general parameters defined by estimating equations. Let the general
population parameter $\xi_{0}$ be defined as the solution to 
\[
\bar{s}_{0}(\xi)=N^{-1}\sum_{i=1}^{N}s(y_{i};\xi)=0,
\]
where $s(y_{i};\xi)$ is the $q$-dimensional estimating function
of $\xi$. For simplicity, we also denote $s(y_{i};\xi)$ by $s_{i}(\xi)$.
These parameters are general and encompass many parameters of interest
in survey sampling. For example, if $s_{i}(\xi)=y_{i}-\xi$, $\xi_{0}$
is the population mean of $y$; if $s_{i}(\xi)=I(y_{i}<c)-\xi$ for
some constant $c$, $\xi_{0}$ is the population proportion of $y$
less than $c$; and if $s_{i}(\xi)=\{y_{i}-\xi_{1},(y_{i}-\xi_{1})^{2}-\xi_{2}\}^{\T}$,
where $\xi=(\xi_{1},\xi_{2})^{\T}$, $\xi_{2,0}$ is the population
variance of $y$. These estimating functions are differentiable with
respect to $\xi$. However, non-differentiable estimating equations
can also be considered. For example, in quantile estimation, let $s_{i}(\xi)=I(y_{i}\leq\xi)-\tau$
for $\tau\in(0,1)$, then $\xi_{0}=\inf\{\xi:\bar{s}_{0}(\xi)\geq0\}$
is the population $100\tau$th quantile. }

\textcolor{black}{We focus on the phase-I and phase-II estimators
(\ref{eq:pIpi}) and (\ref{eq:pIIpi}) of $\bar{s}_{0}(\xi)$ in general
TPRS, denoted as $\bar{s}_{\I}(\xi)$ and $\bar{s}_{\II}(\xi)$, respectively.
Let $\bar{\xi}_{\II}$ be the solution to $\bar{s}_{\II}(\xi)=0$. }

\begin{theorem}\label{Thm-gen-genparameter}\textcolor{black}{Suppose
Assumption \ref{asmp:sampling-pi} holds for $s_{i}(\xi_{0})$ and
that the regularity conditions in Assumption \ref{assumptionB:score}
hold. Under TPRS  in Definition \ref{def:TPRS_gen}, }$\bar{\xi}_{\II}$
follows the limiting distribution:\textcolor{black}{
\[
n_{\II}^{1/2}(\bar{\xi}_{\II}-\xi_{0})\mid\left(Q_{\I}<\gamma^{2}\right)\rightarrow\Gamma_{s}^{\T}(V_{1}^{s})^{1/2}L_{p,\gamma^{2}}+\Gamma_{s}^{\T}(V_{2}^{s})^{1/2}Z_{1}+\Gamma_{s}^{\T}(V_{3}^{s})^{1/2}Z_{2},
\]
as $n_{\II}\rightarrow\infty$, where $\Gamma_{s}=\partial s_{0}(\xi_{0})/\partial\xi$
with $s_{0}(\xi)$ being the limiting function of $\bar{s}_{0}(\xi)$,
\begin{eqnarray}
V_{1}^{s} & = & \lim_{N\rightarrow\infty}n_{\II}B_{0}^{\T}E\left(V_{xx,\I}\mid\F\right)B_{0},\label{eq:V1-s}\\
V_{2}^{s} & = & \lim_{N\rightarrow\infty}n_{\II}E\left(V_{e^{s}e^{s},\I}\mid\F\right),\label{eq:V2-s}\\
V_{3}^{s} & = & \lim_{N\rightarrow\infty}n_{\II}V_{ss,0},\label{eq:V3-s}
\end{eqnarray}
$e_{i}^{s}=s_{i}-(x_{i}^{\T}B_{0})^{\T},$ $B_{0}=\{\sum_{i=1}^{N}(x_{i}-\bar{x}_{0})^{\otimes2}\}^{-1}\sum_{i=1}^{N}(x_{i}-\bar{x}_{0})\{s_{i}(\xi_{0})-\bar{s}_{0}(\xi_{0})\}^{\T}$,
$Z_{1}$ and $Z_{2}$ are standard normal variables, and $(L_{p,\gamma^{2}},Z_{1},Z_{2})$
are jointly independent.}

\end{theorem}

\textcolor{black}{Theorem \ref{Thm-gen-genparameter} includes Theorem
\ref{Thm1-1} as a special case. If $s_{i}(\xi)=y_{i}-\xi$, then
$\Gamma_{s}=1$, and $V_{1}^{s},V_{2}^{s}$ and $V_{3}^{s}$ in (\ref{eq:V1-s})--(\ref{eq:V3-s})
equal $V_{1},V_{2}$ and $V_{3}$ in (\ref{eq:V1})--(\ref{eq:V3}).}

\textcolor{black}{We estimate the asymptotic design variances of $\bar{\xi}_{\II}$
similarly in Section \ref{subsec:ve-pi}. Specifically, the variance
of $\bar{\xi}_{\II}$ can be estimated by $n_{\II}^{-1}\widehat{\Gamma}_{s}^{\T}(\hat{V}_{1}^{s}v_{p,\gamma^{2}}+\hat{V}_{2}^{s}+\hat{V}_{3})\widehat{\Gamma}_{s},$
where $\widehat{\Gamma}_{s}$ is (\ref{eq:pIIpi}) with $u_{i}$$=\partial s(y_{i};\xi)/\partial\xi\mid_{\xi=\bar{\xi}_{\II}}.$
Here, we have 
\begin{eqnarray*}
\hat{V}_{1}^{s} & = & n_{\II}\hat{B}_{\II}^{\T}V_{xx,\I}\hat{B}_{\II},\\
\hat{V}_{2}^{s} & = & \frac{n_{\II}}{N^{2}}\sum_{i\in\cB}\sum_{j\in\cB}\frac{\pi_{\II ij\mid\cA}-\pi_{\II i\mid\cA}\pi_{\II j\mid\cA}}{\pi_{\II i}^{*}\pi_{\II j}^{*}}\frac{(\hat{e}_{i}^{s}-\hat{e}_{\II}^{s})(\hat{e}_{j}^{s}-\hat{e}_{\II}^{s})^{\T}}{\pi_{\II ij\mid\cA}},\\
\hat{V}_{3}^{s} & = & \frac{n_{\II}}{N^{2}}\sum_{i\in\cB}\sum_{j\in\cB}\frac{\pi_{\I ij}-\pi_{\I i}\pi_{\I j}}{\pi_{\I i}\pi_{\I j}}\frac{\{s_{i}-\bar{s}_{\II}(\bar{\xi}_{\II})\}\{s_{j}-\bar{s}_{\II}(\bar{\xi}_{\II})\}^{\T}}{\pi_{\I ij}\pi_{\II ij\mid\cA}},
\end{eqnarray*}
where $s_{i}=s_{i}(\bar{\xi}_{\II})$ for simplicity, $\hat{B}_{\II}=\{\sum_{i=1}^{N}(x_{i}-\bar{x}_{0})^{\otimes2}\}^{-1}\sum_{i=1}^{N}(x_{i}-\bar{x}_{0})\{s_{i}-\bar{s}_{\II}(\bar{\xi}_{\II})\}^{\T}$,
$\hat{e}_{i}^{s}=s_{i}-(x_{i}^{\T}\hat{B}_{\II})^{\T}$, and $\hat{e}_{\II}^{s}$
are (\ref{eq:ybar-II}) with $y_{i}$ being $\hat{e}_{i}^{s}$. Confidence
intervals for components of $\xi_{0}$ can be constructed in a similar
manner as in Section \ref{subsec:ve-pi}.}

\section{Empirical studies\protect\label{sec:Simulation}}

We evaluate the finite-sample performance of rejective sampling through
simulation. We first apply the TPRS with simple random sampling in
Section \ref{subsec:sim1}. We then apply the three-phase rejective
sampling based on an actual study in Section \ref{subsec:sim2} to
showcase the effectiveness of multiple phases.

\subsection{TPRS with simple random sampling\protect\label{subsec:sim1}}

Consider the finite population with bivariate data $\{(x_{i},y_{i}):i=1,\ldots,N=10^{5}\}$,
where $x_{i}\sim0.5^{1/2}(\chi_{1}^{2}-1)$ and $y_{i}=1+\beta x_{i}+e_{i}$
with $e_{i}\sim\N(0,1)$ and $\beta\in\{0.5,1,2\}$. This setup yields
$R^{2}=\beta^{2}/(\beta^{2}+1)$ values of $\{0.2,0.5,0.8\}$.

We implement two-phase simple random sampling with phase-I and phase-II
sample sizes of $n_{\I}=5,000$ and $n_{\II}=200$, respectively,
and the corresponding TPRS with various $\gamma^{2}\in\{0.01,0.05,0.1\}$.
We compare the mean estimators $\bar{y}_{\II}$ and $\bar{y}_{\II,\reg}$
in both sampling scenarios.

We summarize the results in Tables \ref{tab:Sim1-1} and \ref{tab:Sim1-2}.
In all scenarios, $\bar{y}_{\II}$ with rejective sampling improves
the efficiency of $\bar{y}_{\II}$ without rejective sampling with
the percentage of variance reduction increasing with $R^{2}$ and
decreasing with $\gamma^{2}$, which is cohesive with our theoretical
results in Theorem \ref{Thm1} and Corollary \ref{Corollary1}. Table
\ref{tab:Sim1-2} presents the theoretical values of percentage of
variance reduction (\ref{eq:pct=000020varred}) under the simulation
setup. The results in Table \ref{tab:Sim1-1} are consistent with
that in Table \ref{tab:Sim1-2}. The performance of $\bar{y}_{\II}$
with rejective sampling is close to that of $\bar{y}_{\II,\reg}$,
and the results for $\bar{y}_{\II,\reg}$ in both rejective and non-rejective
samples are similar, aligning with our theoretical results in Theorem
\ref{thm:2preg-rej}.\textcolor{black}{{} The variance reduction percentages
for $\bar{y}_{\II,\reg}$ under rejective sampling can be slightly
negative, potentially due to finite sample sizes. }Moreover, the regression
weights in $\bar{y}_{\II,\reg}$ can be negative for $30$ simulated
datasets under two-phase non-rejective sampling; however, the regression
weights are always positive with rejective sampling in this simulation
setting. This demonstrates the practical value of rejective sampling.
Additionally, our variance estimators and confidence intervals, based
on the asymptotic variance formula, prove accurate in both variance
estimation and coverage rates, as evidenced in Table \ref{tab:Sim1-2}.

\begin{table}
\caption{\protect\label{tab:Sim1-1}Simulation results based on $1,000$ Monte
Carlo samples: bias (Bias $\times10^{-2}$), variance (Var $\times10^{-3}$),
mean squared error (MSE $\times10^{-3}$), variance estimate (VE $\times10^{-3}$)
and coverage rate (Cvg $\%$) for $95\%$ confidence intervals calculated
based on the asymptotic variance formula, and the percentage of variance
reduction of the estimator (VarRed $\%$) under rejective sampling
compared to the corresponding estimator under two phase sample random
sampling. $R^{2}=\beta^{2}/(\beta^{2}+1)\in\{0.2,0.5,0.8\}$}

\centering{}%
\begin{tabular}{llcccccc}
 &  &  &  &  &  &  & \tabularnewline
\hline 
 & Method & Bias & Var & MSE & VE & Cvg & {\small{}{}{}VarRed}\tabularnewline
$\gamma^{2}$ &  & $(\times10^{-2})$ & $(\times10^{-3})$ & $(\times10^{-3})$ & $(\times10^{-3})$ & $(\%)$ & $(\%)$\tabularnewline
\hline 
 & \multicolumn{6}{c}{$\beta=0.5$ and $R^{2}=0.2$} & \tabularnewline
\hline 
$\infty$ & $\bar{y}_{\II}$ & -0.30 & 6.0 & 12.1 & 6.2 & 95.5 & --\tabularnewline
 & $\bar{y}_{\II,\reg}$ & -0.22 & 4.9 & 9.9 & 5.0 & 95.3 & --\tabularnewline
\hline 
$0.01$ & $\bar{y}_{\II}$ & 0.35 & 5.0 & 10.0 & 5.0 & 95.4 & 17\tabularnewline
 & $\bar{y}_{\II,\reg}$ & 0.35 & 5.0 & 10.1 & 5.0 & 95.3 & -2\tabularnewline
\hline 
$0.05$ & $\bar{y}_{\II}$ & -0.26 & 5.1 & 10.2 & 5.0 & 94.5 & 16\tabularnewline
 & $\bar{y}_{\II,\reg}$ & -0.27 & 5.1 & 10.2 & 5.0 & 94.3 & -3\tabularnewline
\hline 
$0.1$ & $\bar{y}_{\II}$ & -0.07 & 5.1 & 10.1 & 5.1 & 95.0 & 16\tabularnewline
 & $\bar{y}_{\II,\reg}$ & -0.05 & 5.0 & 10.1 & 5.0 & 95.2 & -2\tabularnewline
\hline 
 & \multicolumn{6}{c}{$\beta=1$ and $R^{2}=0.5$} & \tabularnewline
\hline 
$\infty$ & $\bar{y}_{\II}$ & 0.00 & 10.1 & 20.3 & 10.0 & 94.3 & --\tabularnewline
 & $\bar{y}_{\II,\reg}$ & 0.12 & 5.5 & 11.0 & 5.2 & 94.1 & --\tabularnewline
\hline 
$0.01$ & $\bar{y}_{\II}$ & -0.05 & 5.1 & 10.1 & 5.1 & 94.8 & 50\tabularnewline
 & $\bar{y}_{\II,\reg}$ & -0.05 & 5.0 & 10.1 & 5.2 & 95.1 & 8\tabularnewline
\hline 
$0.05$ & $\bar{y}_{\II}$ & -0.09 & 5.3 & 10.6 & 5.2 & 94.5 & 48\tabularnewline
 & $\bar{y}_{\II,\reg}$ & -0.09 & 5.2 & 10.4 & 5.2 & 94.6 & 5\tabularnewline
\hline 
$0.1$ & $\bar{y}_{\II}$ & -0.17 & 5.3 & 10.7 & 5.3 & 94.5 & 47\tabularnewline
 & $\bar{y}_{\II,\reg}$ & -0.19 & 5.2 & 10.5 & 5.2 & 94.7 & 4\tabularnewline
\hline 
 & \multicolumn{6}{c}{$\beta=2$ and $R^{2}=0.8$} & \tabularnewline
\hline 
$\infty$ & $\bar{y}_{\II}$ & -0.26 & 24.7 & 49.4 & 25.1 & 94.7 & --\tabularnewline
 & $\bar{y}_{\II,\reg}$ & 0.00 & 6.0 & 12.0 & 5.8 & 94.3 & --\tabularnewline
\hline 
$0.01$ & $\bar{y}_{\II}$ & 0.08 & 5.8 & 11.7 & 5.7 & 94.8 & 76\tabularnewline
 & $\bar{y}_{\II,\reg}$ & 0.06 & 5.8 & 11.5 & 5.8 & 94.8 & 4\tabularnewline
\hline 
$0.05$ & $\bar{y}_{\II}$ & -0.35 & 6.0 & 12.1 & 5.9 & 94.9 & 76\tabularnewline
 & $\bar{y}_{\II,\reg}$ & -0.30 & 5.6 & 11.2 & 5.8 & 95.2 & 7\tabularnewline
\hline 
$0.1$ & $\bar{y}_{\II}$ & 0.10 & 6.4 & 12.8 & 6.2 & 94.8 & 74\tabularnewline
 & $\bar{y}_{\II,\reg}$ & 0.09 & 5.9 & 11.7 & 5.8 & 95.0 & 3\tabularnewline
\hline 
\end{tabular}
\end{table}

\begin{table}
\caption{\protect\label{tab:Sim1-2}Theoretical values of percentage of variance
reduction under the simulation setup}

\centering{}%
\begin{tabular}{llccc}
 &  &  &  & \tabularnewline
\hline 
 & $\gamma^{2}$ & 0.01 & 0.05 & 0.1\tabularnewline
 & $v_{p,\gamma^{2}}$ & 0.003 & 0.017 & 0.033\tabularnewline
\hline 
$\frac{1-f_{\II,\I}}{1-f_{\II,\I}f_{\I,\F}}\left(1-v_{p,\gamma^{2}}\right)R^{2}$ & $R^{2}=0.2$ & 19.2 & 47.9 & 76.7\tabularnewline
 & $R^{2}=0.5$ & 18.9 & 47.3 & 75.7\tabularnewline
 & $R^{2}=0.8$ & 18.6 & 46.5 & 74.4\tabularnewline
\hline 
\end{tabular}
\end{table}

\subsection{Three-phase sampling based on the Academic Performance Index data\protect\label{subsec:sim2} }

To demonstrate the practical relevance, we use the Academic Performance
Index (API) data. The full population data consist of $6,194$ observations
for all California schools with at least 100 students based on standardized
testing of students. We use the API in 2000 as the study variable
$y$, the API in 1999 as the auxiliary variable $x$, and the percentage
of English Language Learners, the percentage of students eligible
for subsidized meals, the percentage of students for whom this is
the first year at the school as additional auxiliary variable $z$
in both the design and analysis stages. The parameter of interest
is the population mean of the API in 2000.

We employed three-phase sampling designs. We select a phase-I sample
$\cA$ of size $n_{\I}=2000$ by simple random sampling. For $i\in\cA$,
we observe $x_{i}$. We select a phase-II sample $\cB$ of size $n_{\II}$
by Poisson sampling with $\pi_{\II i}=p_{\II i}E(n_{\II})$, where
$E(n_{\II})$ is the expected sample size of the phase-II sample,
$p_{\II i}\propto x_{i}$ and $\sum_{i\in\cA}p_{\II i}=1$. For $i\in\cB$,
we observe $z_{i}$ and calculate $a_{i}=z_{i}-\bar{z}_{\II}-(x_{i}-\bar{x}_{\II})^{\T}\hat{\beta}_{zx,\II}$.
We select a phase-$\III$ sample $\cC$ by Poisson sampling with $\pi_{\III i}=p_{\III i}E(n_{\III}),$
where $E(n_{\III})$ is the expected sample size of the phase-III
sample, $p_{\III i}$ is proportional to the summation of components
in $z_{i}$ and $\sum_{i\in\cB}p_{\III i}=1$. For $i\in\cC$, we
observe $y_{i}$. We consider the above three-phase sampling without
rejective sampling and with rejective sampling described in Section
\ref{subsec:Three-phase-rejective-sampling}. \textcolor{black}{We
consider the impact of the number of phases, phase sample sizes $E(n_{\II})$
and $E(n_{\III})$, and constraints $\gamma_{1}^{2}$ and $\gamma_{2}^{2}$.}\textcolor{blue}{{}
}For comparison, we also consider two-phase sampling designs. In a
two-phase sampling, we assume that the outcome is measured for the
phase-$\II$ samples. This may be more costly than the three-phase
sampling.

\begin{figure}
\centering

\includegraphics[scale=0.7]{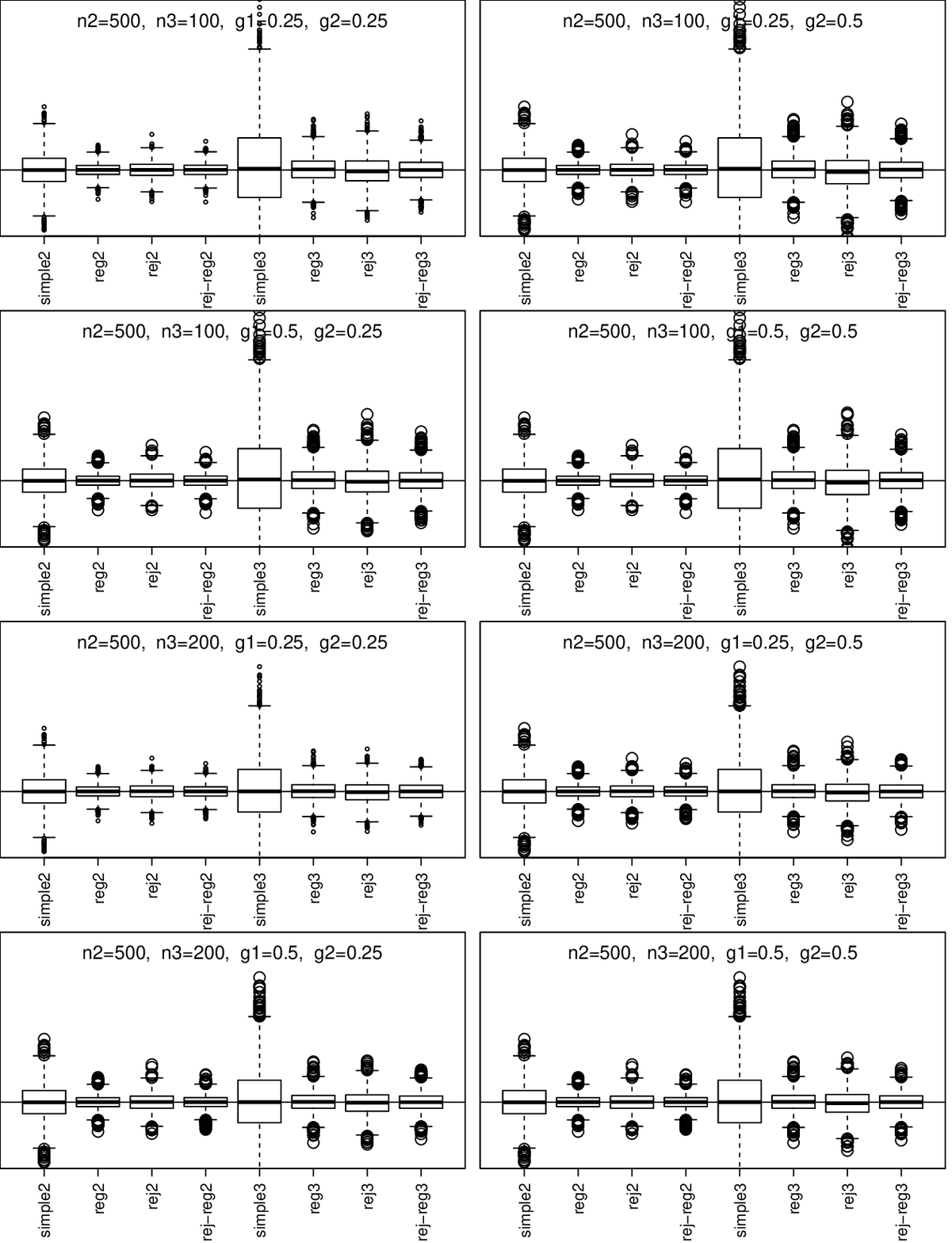}

\vspace{0.25cm}

{\footnotesize Footnote: n2 is $E(n_{\II})$, n3 is $E(n_{\III})$,
g1 is $\gamma_{1}^{2}$, and g2 is $\gamma_{2}^{2}$; simple2 and
reg2 are $\bar{y}_{\II}$ and $\bar{y}_{\II,\reg}$ without rejective
sampling; rej2 and rej-reg2 are $\bar{y}_{\II}$ and $\bar{y}_{\II,\reg}$
with rejective sampling; simple3 and reg3 are $\bar{y}_{\III}$ and
$\bar{y}_{\III,\reg}$ without rejective sampling; rej3 and rej-reg3
are $\bar{y}_{\III}$ and $\bar{y}_{\III,\reg}$ with rejective sampling.}{\footnotesize\par}

\caption{\protect\label{fig:2-1}Simulation results for simple estimators and
regression estimators under two/three-phase w/o rejective sampling
with $n_{\II}=500$}
\end{figure}

\begin{figure}
\centering

\includegraphics[scale=0.7]{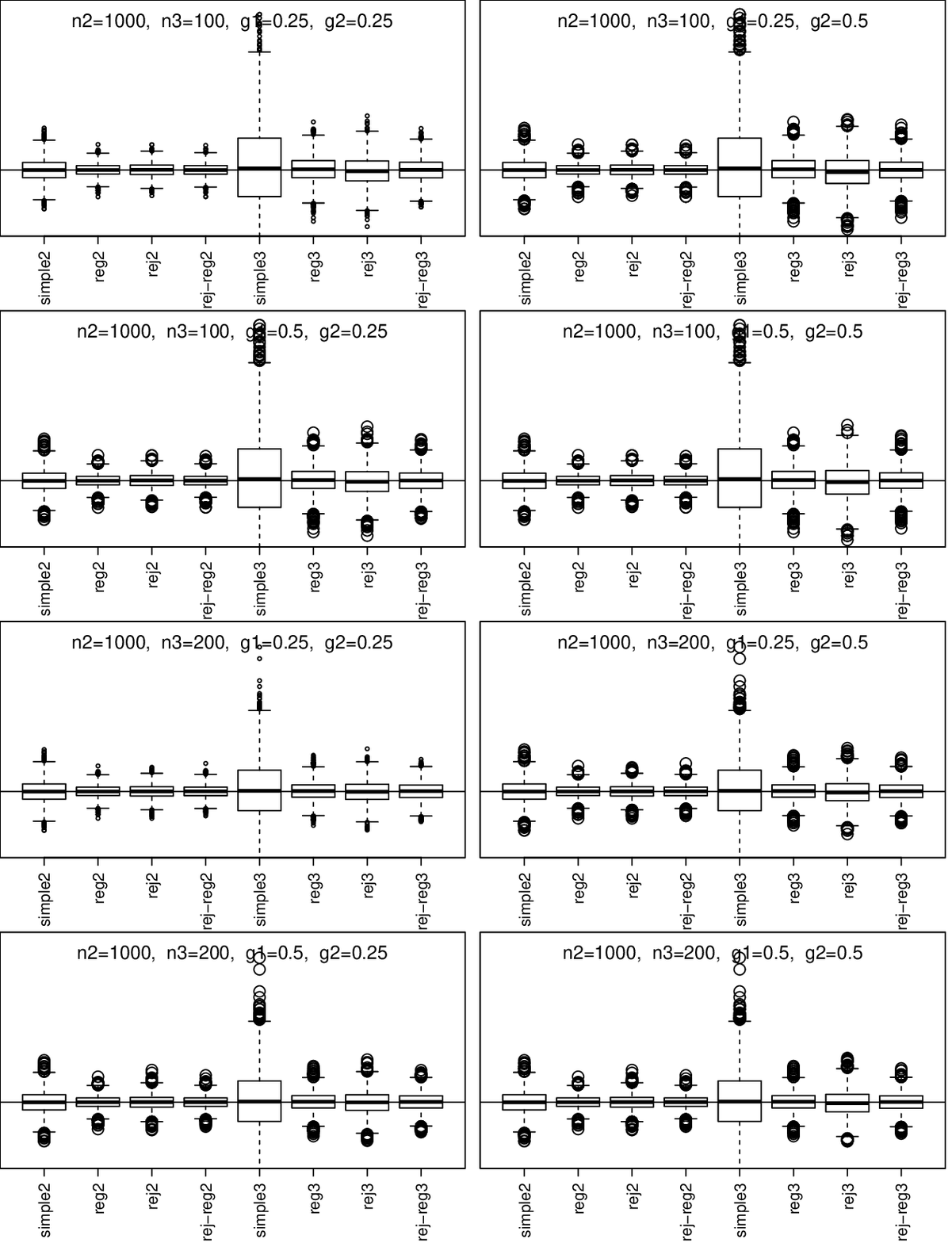}

\vspace{0.25cm}

{\footnotesize Footnote: n2 is $E(n_{\II})$, n3 is $E(n_{\III})$,
g1 is $\gamma_{1}^{2}$, and g2 is $\gamma_{2}^{2}$; simple2 and
reg2 are $\bar{y}_{\II}$ and $\bar{y}_{\II,\reg}$ without rejective
sampling; rej2 and rej-reg2 are $\bar{y}_{\II}$ and $\bar{y}_{\II,\reg}$
with rejective sampling; simple3 and reg3 are $\bar{y}_{\III}$ and
$\bar{y}_{\III,\reg}$ without rejective sampling; rej3 and rej-reg3
are $\bar{y}_{\III}$ and $\bar{y}_{\III,\reg}$ with rejective sampling.}{\footnotesize\par}

\caption{\protect\label{fig:2-2}Simulation results for simple estimators and
regression estimators under two/three-phase w/o rejective sampling
with $n_{\II}=1000$}
\end{figure}

Simulation results, summarized in Figures \ref{fig:2-1} and \ref{fig:2-2},
show that phase-II estimators are more efficient due to more information
collection, but phase-III designs might provide greater cost-effectiveness.
For instance, phase-II rejective regression estimators with an average
sample size of $n_{\II}=500$ had a variance of $5.4$, compared with
phase-III estimators with a smaller average sample size of $n_{\III}=100$
but a higher variance of $15.9$. This implies that multiple phases
can achieve desired variances with smaller samples. The efficiency
of rejective regression estimators improves with tighter constraints
$\gamma_{1}^{2}$ and $\gamma_{2}^{2}$ , especially $\gamma_{2}^{2}$
. Increasing the phase-II sample size showed minimal impact on the
efficiency of phase-III estimator. The coverage rates of the confidence
intervals align well with the nominal level, confirming our results
on the asymptotic distributions.

\begin{figure}
\centering{}\includegraphics[height=10cm]{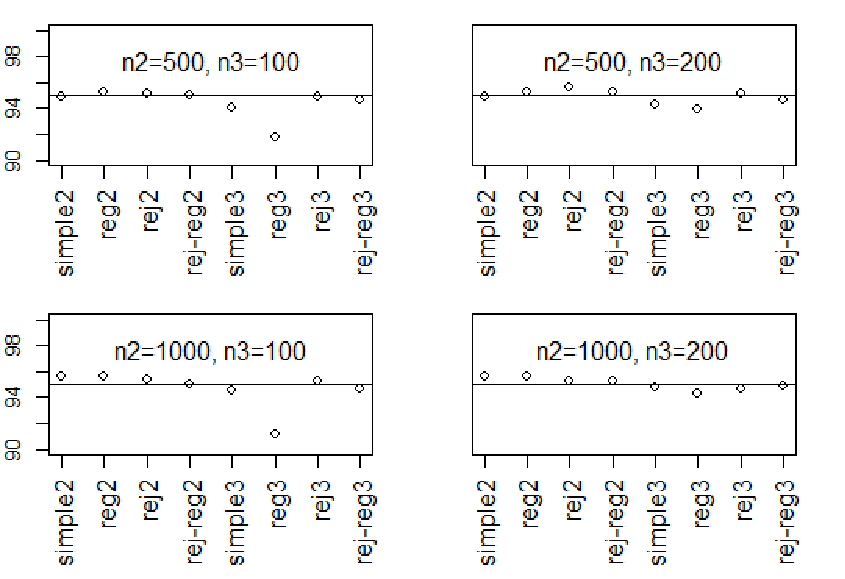}

\vspace{0.25cm}

{\footnotesize Footnote: n2 is $E(n_{\II})$, n3 is $E(n_{\III})$,
simple2 and reg2 are $\bar{y}_{\II}$ and $\bar{y}_{\II,\reg}$ without
rejective sampling; rej2 and rej-reg2 are $\bar{y}_{\II}$ and $\bar{y}_{\II,\reg}$
with rejective sampling; simple3 and reg3 are $\bar{y}_{\III}$ and
$\bar{y}_{\III,\reg}$ without rejective sampling; rej3 and rej-reg3
are $\bar{y}_{\III}$ and $\bar{y}_{\III,\reg}$ with rejective sampling.}{\footnotesize\par}

\caption{\protect\label{fig:2-3-cvg}Coverage rates}
\end{figure}

\section*{Acknowledgments}

We thank the associate editor and two reviewers for helpful comments.
Chongguang Tao proofread the final version of the paper. Yang was
partially funded by the U.S. National Institutes of Health (grant
\# 1R01AG066883) and the U.S. National Science Foundation (grant \#
SES 2242776). Ding was partially funded by the U.S. National Science
Foundation (grant \# 1945136). 

\bibliographystyle{dcu}
\bibliography{/Users/syang24/Dropbox/bib/cinew,/Users/syang24/Dropbox/bib/VEriverbmk_pfi_MIsurvey}

\newpage{}

\global\long\def\theequation{S\arabic{equation}}%
\setcounter{equation}{0}

\global\long\def\thelemma{S\arabic{lemma}}%
\setcounter{lemma}{0}

\global\long\def\thetable{S\arabic{table}}%
\setcounter{table}{0}

\global\long\def\theexample{S\arabic{example}}%
\setcounter{equation}{0}

\global\long\def\thedefinition{S\arabic{definition}}%
\setcounter{definition}{0}

\global\long\def\thesection{S\arabic{section}}%
\setcounter{section}{0}

\global\long\def\thetheorem{S\arabic{theorem}}%
\setcounter{equation}{0}

\global\long\def\thecondition{S\arabic{condition}}%
\setcounter{equation}{0}

\global\long\def\theremark{S\arabic{remark}}%
\setcounter{equation}{0}

\global\long\def\thestep{S\arabic{step}}%
\setcounter{equation}{0}

\global\long\def\theassumption{S\arabic{assumption}}%
\setcounter{assumption}{0}

\global\long\def\theproof{S\arabic{proof}}%
\setcounter{equation}{0}

\global\long\def\theproposition{S{proposition}}%
\setcounter{equation}{0}

\global\long\def\thefigure{S\arabic{figure}}%
\setcounter{figure}{0} 

\global\long\def\thepage{S\arabic{page}}%
\setcounter{page}{0}
\begin{center}
{\huge\textbf{Supplementary material}}
\par\end{center}

\setcounter{page}{1} \pagenumbering{arabic}

\pagestyle{fancy}
\renewcommand{\thepage}{S\arabic{page}}

Section \ref{sec:Sequential-rejective-sampling} describes sequential
rejective sampling with tiers of covariates. Section \ref{sec:Multi-phase-rejective-sampling}
presents multi-phase rejective sampling. Section \ref{sec:Proofs}
provides all proofs. 

\section{Sequential rejective sampling with tiers of covariates\protect\label{sec:Sequential-rejective-sampling}}

Partition covariate $x$ into $K$ tiers of decreasing importance,
denoted by $(x[1],\ldots,x[K])$, each $x[k]$ having dimension $p[k]$.
Let $x[\overline{k}]=(x[1],\ldots,x[k])$. It is convenient to apply
the block-wise Gram--Schmidt orthogonalization to the covariates:
construct $g=(g[1],\ldots,g[K])$ from $g[1]=x[1]$, and 
\[
g[k]=x[k]-V_{x[k]x[\overline{k-1}],\I}\left(V_{x[\overline{k-1}]x[\overline{k-1}],\I}\right)^{-1}x[\overline{k-1}]
\]
for $k=2,\ldots,K$. We assume $\cov(\bar{g}_{\II}[k],\bar{g}_{\II}[l]\mid\cA)=o_{P}(n_{\II}^{-1})$
$\as$ for $k\neq l$ under suitable regularity conditions. See the
discussion of Assumption \ref{asmp:sampling-pi}(vii) in the main
paper. For each tier $k$, we can impose a threshold $\omega_{k}^{-1}\gamma^{2}$,
where $\gamma^{2}$ is a positive constant and $\omega_{k}$ is the
weight for tier $k$, with higher $\omega_{k}$ indicating a stricter
threshold.

We define two-phase sequential rejective sampling (TPSRS) as follows.

\begin{definition}[TPSRS with general sampling]\label{def:TPRS_seq}
TPSRS with general sampling consists of two steps: 
\begin{description}
\item [{Step$\ 1.$}] Select a phase-I sample $\cA$ by a general $\pi$
sampling with the inclusion probability $\pi_{\I i}$. For $i\in\cA,$
record $x_{i}$. 
\item [{Step$\ 2.$}] Treat the phase-I sample $\cA$ as the population
and select a phase-II sample $\cB$ by a general $\pi$ sampling with
the conditional inclusion probability $\pi_{\II i\mid\cA}$\textcolor{red}{{}
}given unit $i$ is in the phase-$\I$ sample. Accept the phase-II
sample if
\begin{equation}
Q_{\I}[k]=(\bar{g}_{\II}[k]-\bar{g}_{\I}[k])^{\T}V_{g[k]g[k],\I}^{-1}(\bar{g}_{\II}[k]-\bar{g}_{\I}[k])<\omega_{k}^{-1}\gamma^{2},\ (1\leq k\leq K),\label{eq:tierk}
\end{equation}
done sequentially from tier $1$ to $K$. For $i\in\cB,$ record $y_{i}$. 
\end{description}
\end{definition}

\begin{remark}In a weighted rejective sampling, we accept the phase-II
sample if
\begin{equation}
\sum_{k=1}^{K}\omega_{k}Q_{\I}[k]<K\gamma^{2},\label{eq:tierk-weighted}
\end{equation}
where the $\omega_{k}$'s may be different from the ones in the sequential
rejective sampling as in (\ref{eq:tierk}).

Sequential rejective sampling, unlike weighted rejective sampling,
ensures balance across all tiers, as it requires each tier to individually
meet its threshold. Weighted rejective sampling is less stringent,
allowing some tiers to be poorly balanced if others are well balanced.
\textcolor{black}{\citet{zhong2024conditionally} studied the admissibility
of different rerandomization criteria.} The limiting distributions
of $\bar{y}_{\II}$ and $\bar{y}_{\II,\reg}$ under weighted rejective
sampling (\ref{eq:tierk-weighted}) are more complicated compared
with the TPSRS in Definition \ref{def:TPRS_seq} \citep{lu2023design}.

\end{remark}

To study the limiting properties of $\bar{y}_{\II}$, let $\beta_{gy}$
be the probability limit of $V_{gg,\I}^{-1}V_{gy,\I}$. By the Gram--Schmidt
orthogonalization, $V_{gg,\I}$ is a diagonal matrix with the $k$th
block diagonal component $V_{g[k]g[k],\I}$. We decompose $\beta_{gy}$
as $\beta_{gy}=(\beta_{g[1]y},\ldots,\beta_{g[K]y})$ with $\beta_{g[k]y}$
being the probability of $V_{g[k]g[k],\I}^{-1}V_{g[k]y,\I}$. The
decomposition of $\bar{y}_{\II}-\bar{y}_{0}$ then becomes 
\begin{eqnarray*}
\bar{y}_{\II}-\bar{y}_{0} & = & (\bar{g}_{\II}-\bar{g}_{\I})^{\T}\beta_{gy}+\left\{ \bar{y}_{\II}-\bar{y}_{\I}-(\bar{g}_{\II}-\bar{g}_{\I})^{\T}\beta_{gy}\right\} +\bar{y}_{\I}-\bar{y}_{0}\\
 & = & \sum_{k=1}^{K}(\bar{g}_{\II}[k]-\bar{g}_{\I}[k])^{\T}\beta_{g[k]y}+\left\{ \bar{y}_{\II}-\bar{y}_{\I}-(\bar{g}_{\II}-\bar{g}_{\I})^{\T}\beta_{gy}\right\} +\bar{y}_{\I}-\bar{y}_{0}.
\end{eqnarray*}
Let $D_{\I}[k]=V_{g[k]g[k],\I}^{-1/2}(\bar{g}_{\II}[k]-\bar{g}_{\I}[k])$,
then $D_{\I}[k]\rightarrow\N(0,I_{p[k]})$. For any $k\neq l$, $\bar{g}_{\II}[k]-\bar{g}_{\I}[k]$
and $\bar{g}_{\II}[l]-\bar{g}_{\I}[l]$ are asymptotically independent
given $\cA$ $\as$. Therefore, 
\[
\sum_{k=1}^{K}(\bar{g}_{\II}[k]-\bar{g}_{\I}[k])^{\T}\beta_{g[k]y}\mid\left(D_{\I}^{\T}[k]D_{\I}[k]<\omega_{k}^{-1}\gamma^{2}:1\leq k\leq K\right)
\]
has the same limiting distribution of 
\[
\sum_{k=1}^{K}(\bar{g}_{\II}[k]-\bar{g}_{\I}[k])^{\T}\beta_{g[k]y}\mid\left(D_{\I}^{\T}[k]D_{\I}[k]<\omega_{k}^{-1}\gamma^{2}\right).
\]
Theorem \ref{Thm1-1-1} below describes the limiting distributions
of $\bar{y}_{\II}$ and $\bar{y}_{\II,\reg}$ under TPSRS with general
sampling.

\begin{theorem}\label{Thm1-1-1}Suppose Assumption \ref{asmp:sampling-pi}
holds. Under TPSRS in Definition \ref{def:TPRS_seq}, $\bar{y}_{\II}$
follows the limiting distribution:
\begin{equation}
n_{\II}^{1/2}(\bar{y}_{\II}-\bar{y})\mid\left(D_{\I}^{\T}[k]D_{\I}[k]<\omega_{k}^{-1}\gamma^{2}:1\leq k\leq K\right)\rightarrow\sum_{k=1}^{K}V_{1}^{1/2}[k]L_{p[k],\omega_{k}^{-1}\gamma^{2}}+V_{2}^{1/2}Z_{1}+V_{3}^{1/2}Z_{2},\label{eq:rej-normal-1-1}
\end{equation}
 where $V_{1}[k]=\lim_{n_{\II}\rightarrow\infty}n_{\II}\beta_{g[k]y}^{\T}E(V_{g[k]g[k],\I}\mid\F)\beta_{g[k]y},$
$V_{2}$ and $V_{3}$ are defined in (\ref{eq:V2}) and (\ref{eq:V3}),
respectively, $Z_{1}$ and $Z_{2}$ are independent standard normal
variables, and $(L_{p[k],\omega_{k}^{-1}\gamma^{2}},Z_{1},Z_{2})$
are jointly independent. Moreover, 
\[
n_{\II}^{1/2}(\bar{y}_{\II,\reg}-\bar{y}_{0})\mid\left(D_{\I}^{\T}[k]D_{\I}[k]<\omega_{k}^{-1}\gamma^{2}:1\leq k\leq K\right)\rightarrow V_{2}^{1/2}Z_{1}+V_{3}^{1/2}Z_{2}.
\]

\end{theorem}

\section{Multi-phase rejective sampling\protect\label{sec:Multi-phase-rejective-sampling}}

\subsection{Notation}

We extend our theory to multi-phase rejective sampling. To simplify
the presentation, we discuss three-phase sampling below. Let $\pi_{\I i}$,
$\pi_{\II i\mid\cA}$, and $\pi_{\III i\mid\cB}$ represent the probabilities
of a unit being included in the phase-I, phase-II, and phase-III samples,
with each phase indexed by $\cA$, $\cB$ and $\cC$, respectively.
The combined probability for phase-III is $\pi_{\III i}^{*}=\pi_{\I i}\pi_{\II i\mid\cA}\pi_{\III i\mid\cB}$.
Let $n_{\I}$, $n_{\II}$, and $n_{\III}$ be the sampling sizes of
the phase-I sample, the phase-II sample, and the phase-III sample,
respectively. 

For the finite population mean $\bar{u}_{0}$, the phase-$\I$, phase-II,
and phase-III estimators are (\ref{eq:pIpi}), (\ref{eq:pIIpi}),
and 
\begin{equation}
\bar{u}_{\III}=\frac{1}{\sum_{i\in\cC}(\pi_{\III i}^{*})^{-1}}\sum_{i\in\cC}\frac{u_{i}}{\pi_{\III i}^{*}},\label{eq:pIIIpi}
\end{equation}
respectively. The design variance for the phase-III estimator involves
conditional second-order inclusion probabilities $\pi_{\III ij\mid\cB}$.
By Taylor expansion and ignoring the small order terms, the design
covariance of $\bar{u}_{\III\pi}$ and $\bar{v}_{\III\pi}$ given
the phase-$\II$ sample is 
\begin{equation}
V_{uv,\II}=\cov\left(\bar{u}_{\III\pi},\bar{v}_{\III\pi}\mid\cB,\cA,\F\right)=\frac{1}{N^{2}}\sum_{i\in\cB}\sum_{j\in\cB}\frac{\pi_{\III ij\mid\cB}-\pi_{\III i\mid\cB}\pi_{\III j\mid\cB}}{\pi_{\III i}^{*}\pi_{\III j}^{*}}(u_{i}-\bar{u}_{\II})(v_{j}-\bar{v}_{\II})^{\T}.\label{eq:V_uv2-pi}
\end{equation}

\subsection{Three-phase rejective sampling\protect\label{subsec:Three-phase-rejective-sampling}}

We define a three-phase rejective sampling method with $p_{1}$-dimensional
phase-I covariate $x$ and $p_{2}$-dimensional phase-II covariate
$z$ as follows.

\begin{definition}[Three-phase rejective sampling with general sampling]\label{def:ThreePRS}
Three-phase rejective sampling with general sampling consists of three
steps: 
\begin{description}
\item [{Step$\ 1.$}] Select a phase-$\I$ sample $\cA$ by a general $\pi$
sampling with the inclusion probability $\pi_{\I i}$. For $i\in\cA,$
record $x_{i}$. 
\item [{Step$\ 2.$}] Select a phase-$\II$ sample $\cB$ by a general
$\pi$ sampling with the conditional inclusion probability $\pi_{\II i\mid\cA}$
given unit $i$ is in the phase-$\I$ sample. Accept the phase-$\II$
sample if 
\[
Q_{x,\I}=\left(\bar{x}_{\II}-\bar{x}_{\I}\right)^{\T}V_{xx,\I}^{-1}\left(\bar{x}_{\II}-\bar{x}_{\I}\right)<\gamma_{1}^{2},
\]
where $\gamma_{1}^{2}>0$, and $V_{xx,\I}$ is the positive definite
design variance of $\bar{x}_{\II}-\bar{x}_{\I}$ given the phase-$\I$
sample. For $i\in\cB,$ record $z_{i}$. It is convenient to apply
the Gram--Schmidt orthogonalization to $z_{i},$ and construct a
phase-$\II$ covariate $a_{i}=z_{i}-\bar{z}_{\II}-(x_{i}-\bar{x}_{\II})^{\T}\hat{\beta}_{zx,\II}$,
where 
\begin{equation}
\hat{\beta}_{zx,\II}=\left\{ \sum_{i\in\cB}\frac{(x_{i}-\bar{x}_{\II})^{\otimes2}}{\pi_{\II i}^{*}}\right\} ^{-1}\sum_{i\in\cB}\frac{(x_{i}-\bar{x}_{\II})(z_{i}-\bar{z}_{\II})^{\T}}{\pi_{\II i}^{*}}.\label{eq:hat-betaII}
\end{equation}
Let $c=(x^{\T},a^{\T})^{\T}$ represent the combined vector of phase-$\II$
covariates, where $a$ can be seen as the part of $z$ that is both
correlated with $y$ and orthogonal to $x$, representing the additional
information beyond what is available from phase $\I$. 
\item [{Step$\ 3.$}] Select a phase-$\III$ rejective sample $\cC$ by
a general $\pi$ sampling with the conditional inclusion probability
$\pi_{\III i\mid\cB}$ given unit $i$ is in the phase-$\II$ sample.
Accept the phase-$\III$ sample if 
\[
Q_{c,\II}=\left(\bar{c}_{\III}-\bar{c}_{\II}\right)^{\T}V_{cc,\II}^{-1}\left(\bar{c}_{\III}-\bar{c}_{\II}\right)<\gamma_{2}^{2},
\]
where $\gamma_{2}^{2}>0$, and $V_{cc,\II}$ is the positive definite
design variance of $\bar{c}_{\III}-\bar{c}_{\II}$ given the phase-$\II$
sample. For $i\in\cC,$ record  $y_{i}$. 
\end{description}
\end{definition}

In Step 3, balance can be directly controlled on $(x^{\T},z^{\T})^{\T}$
instead of $c$, because $\bar{c}_{\III}-\bar{c}_{\II}$ is a linear
transformation of $\bar{x}_{\III}-\bar{x}_{\II}$ and $\bar{z}_{\III}-\bar{z}_{\II}$
given $\hat{\beta}_{zx,\II}$ and the Mahalanobis distance is invariant
to linear transformations.

For the population mean $\bar{y}_{0}$, the $\pi^{*}$ estimator based
on the phase-$\III$ rejective sample is 
\[
\bar{y}_{\III}=\frac{1}{\sum_{i\in\cC}(\pi_{\III i}^{*})^{-1}}\sum_{i\in\cC}\frac{y_{i}}{\pi_{\III i}^{*}}.
\]
To derive the asymptotic properties of $\bar{y}_{\III}$, we need
the following regularity condition. 

\begin{assumption}\label{asmp:sampling-pi-3phase} The phase-$\III$
estimator (\ref{eq:pIIIpi}) satisfies 
\[
\var(\hat{u}_{\III}\mid\cB,\cA,\F)^{-1/2}(\hat{u}_{\III}-\bar{u}_{\II})\mid\cB,\cA,\F\rightarrow\N(0,1)\ \as,
\]
with $\var(n_{\III}^{1/2}\hat{u}_{\III}\mid\cB,\cA,\F)=O_{P}(1)$,
where $u$ represents components of either $x$ or $y$. 

The sequence of phase-$\II$ selection probabilities are bounded by
$K_{\III,\textup{L}}<n_{\III}^{-1}n_{\II}\pi_{\III i}<K_{\III,\textup{U}}$
for all $i,$ for some positive $K_{\III,\textup{L}}>0$ and $K_{\III,\textup{U}}>0$.
Assume that the design is such that 
\[
\lim_{N\rightarrow\infty}\sum_{i\in\cC}\pi_{\III i}^{-1}(1,c_{i}^{\T},y_{i},y_{i}^{2})^{\T}(1,c_{i}^{\T},y_{i},y_{i}^{2})=M_{\III}\ \as,
\]
where $M_{\III}$ is a matrix of constants. Moreover, $\cov(\bar{c}_{\III}-\bar{c}_{\II},\bar{e}_{\III}-\bar{e}_{\II}\mid\cB,\cA,\F)=o_{P}(n_{\III}^{-1})$.

\end{assumption}

\begin{theorem}\label{Thm:3}Suppose Assumptions \ref{asmp:sampling-pi}
and \ref{asmp:sampling-pi-3phase} hold. Under three-phase rejective
sampling in Definition \ref{def:ThreePRS}, $\bar{y}_{\III}$ follows
the limiting distribution:
\begin{multline}
n_{\III}^{1/2}\left(\bar{y}_{\III}-\bar{y}_{0}\right)\mid(Q_{z,\I}<\gamma_{1}^{2},Q_{c,\II}<\gamma_{2}^{2})\\
\rightarrow V_{\III,1}^{1/2}L_{p_{1}+p_{2},\gamma_{2}^{2}}+V_{\III,2}^{1/2}Z_{1}+V_{\III,3}^{1/2}L_{p_{1},\gamma_{1}^{2}}+V_{\III,4}^{1/2}Z_{2}+V_{\III,5}^{1/2}Z_{3},\label{eq:yIII-dist}
\end{multline}
where 
\begin{align}
V_{\III,1} & =\lim_{N\rightarrow\infty}n_{\III}\beta_{yc,0}^{\T}E\left(V_{cc,\II}\mid\F\right)\beta_{yc,0}, & V_{\III,2} & =\lim_{N\rightarrow\infty}n_{\III}E\left(V_{e_{yc}e_{yc},\II}\mid\F\right),\nonumber \\
V_{\III,3} & =\lim_{N\rightarrow\infty}n_{\III}\beta_{yx,0}^{\T}E(V_{xx,\I}\mid\F)\beta_{yx,0}, & V_{\III,4} & =\lim_{N\rightarrow\infty}n_{\III}E\left(V_{e_{yx}e_{yx},\I}\mid\F\right),\nonumber \\
V_{\III,5} & =\lim_{N\rightarrow\infty}n_{\III}V_{yy,0}, & e_{uv} & =u-v^{\T}\beta_{uv,0},\label{eq:e-uv}
\end{align}
$Z_{1}$, $Z_{2}$, and $Z_{3}$ are standard normal variables, and
$(Z_{1},Z_{2},Z_{3},L_{p_{1},\gamma_{1}^{2}},L_{p_{1}+p_{2},\gamma_{2}^{2}})$
are jointly independent.

\end{theorem}

\subsection{Regression estimator\protect\label{subsec:Regression-estimator}}

Integrating the design and analysis strategies, the three-phase regression
estimator of $\bar{y}_{0}$ is 
\begin{equation}
\bar{y}_{\III,\reg}=\bar{y}_{\III}+\left(\begin{array}{c}
\bar{x}_{\I}-\bar{x}_{\III}\\
\bar{a}_{\II}-\bar{a}_{\III}
\end{array}\right)^{\T}\hat{\beta}_{yc,\III},\label{eq:IIIreg}
\end{equation}
where $\bar{a}_{\II}=0$ by our construction of $a_{i}$, and

\begin{equation}
\hat{\beta}_{yc,\III}=\left\{ \sum_{i\in\cC}\frac{\left(c_{i}-\bar{c}_{\III}\right)\left(c_{i}-\bar{c}_{\III}\right)^{\T}}{\pi_{\III i}^{*}}\right\} ^{-1}\sum_{i\in\cC}\frac{(c_{i}-\bar{c}_{\III})(y_{i}-\bar{y}_{\III})}{\pi_{\III i}^{*}}.\label{eq:beta-hatycIII}
\end{equation}
\textcolor{blue}{}

The regression estimator exhibits\textcolor{red}{{} }the same asymptotic
behavior, irrespective of the use of rejective sampling.

\begin{theorem}\label{Thm3-reg}Under Assumptions \ref{asmp:sampling-pi}
and \ref{asmp:sampling-pi-3phase}, under three-phase rejective sampling
in Definition \ref{def:ThreePRS}, the regression estimator $\bar{y}_{\III,\reg}$
in (\ref{eq:IIIreg}) has the following limiting distribution:
\[
n_{\III}^{1/2}(\bar{y}_{\III,\reg}-\bar{y}_{0})\mid\F\rightarrow\N\left(0,V_{\III,\reg}\right),
\]
$\as$ for all sequences of finite populations, where 
\[
V_{\III,\reg}=\lim_{N\rightarrow\infty}n_{\III}\left\{ V_{yy,0}+E\left(V_{e_{yx}e_{yx},\I}\mid\F\right)+E\left(V_{e_{yc}e_{yc},\II}\mid\F\right)\right\} ,
\]
and $V_{uv,0}$, $V_{uv,\I}$ and $V_{uv,\II}$ are defined in (\ref{eq:V_uvN-pi}),
(\ref{eq:V_uv1-pi}) and (\ref{eq:V_uv2-pi}), respectively.

\end{theorem}

Thus, the limiting distribution of $n_{\III}^{1/2}(\bar{y}_{\III,\reg}-\bar{y}_{0})$
remains unchanged with or without rejective procedure. Combining the
results from Theorems \ref{Thm:3} and \ref{Thm3-reg} suggests that
utilizing auxiliary variables in both the design stage (via rejective
sampling) and analysis stage (via regression) can improve estimation
efficiency.

\subsection{Inference: variance estimators and confidence intervals\protect\label{subsec:ve-pi-1}}

We derive the asymptotic design variance formula. Let $\hat{\beta}_{uv,\III}$
be (\ref{eq:beta-hatycIII}) with $y_{i}$ and $c_{i}$ being $u_{i}$
and $v_{i}$, $\hat{e}_{uv,i}=u_{i}-v_{i}^{\T}\hat{\beta}_{uv,\III}$,
and $\hat{e}_{uv,\III}$ be (\ref{eq:pIIIpi}) with $u_{i}$ being
$\hat{e}_{uv,i}$. We estimate $V_{yy,0}$, $V_{e_{yx}e_{yx},\I}$
and $V_{e_{yc}e_{yc},\II}$ by 
\begin{eqnarray*}
\hat{V}_{yy,0} & = & \frac{1}{N^{2}}\sum_{i\in\cA_{\III}}\sum_{j\in\cA_{\III}}\frac{\pi_{\I ij}-\pi_{\I i}\pi_{\I j}}{\pi_{\I i}\pi_{\I j}}\frac{(y_{i}-\bar{y}_{\III})(y_{j}-\bar{y}_{\III})^{\T}}{\pi_{\I ij}\pi_{\II ij\mid\cA}\pi_{\III ij\mid\cB}},\\
\hat{V}_{e_{yx}e_{yx},\I} & = & \frac{1}{N^{2}}\sum_{i\in\cA_{\III}}\sum_{j\in\cA_{\III}}\frac{\pi_{\II ij\mid\cA}-\pi_{\II i\mid\cA}\pi_{\II j\mid\cA}}{\pi_{\II i}^{*}\pi_{\II j}^{*}}\frac{(\hat{e}_{yx,i}-\hat{e}_{yx,\III})(\hat{e}_{yx,j}-\hat{e}_{yx,\III})^{\T}}{\pi_{\II ij\mid\cA}\pi_{\III ij\mid\cA}},\\
\hat{V}_{e_{yc}e_{yc},\II} & = & \frac{1}{N^{2}}\sum_{i\in\cA_{\III}}\sum_{j\in\cA_{\III}}\frac{\pi_{\III ij\mid\cB}-\pi_{\III i\mid\cB}\pi_{\III j\mid\cB}}{\pi_{\III i}^{*}\pi_{\III j}^{*}}\frac{(\hat{e}_{yc,i}-\hat{e}_{yc,\III})(\hat{e}_{yc,j}-\hat{e}_{yc,\III})^{\T}}{\pi_{\III ij\mid\cB}},
\end{eqnarray*}
respectively. Then, the variance estimator for $\bar{y}_{\III}$ is
\[
\hat{V}_{\III}=\left(\hat{\beta}_{yc,\III}^{\T}V_{cc,\II}\hat{\beta}_{yc,\III}\right)v_{p_{1}+p_{2},\gamma_{2}^{2}}+\left(\hat{\beta}_{yx,\III}^{\T}V_{xx,\I}\hat{\beta}_{yx,\III}\right)v_{p_{1},\gamma_{1}^{2}}+\hat{V}_{e_{yc}e_{yc},\II}+\hat{V}_{e_{yx}e_{yx},\I}+\hat{V}_{yy,0}.
\]
The variance estimator for $\bar{y}_{\III,\reg}$ is $\hat{V}_{\III,\reg}=\hat{V}_{yy,0}+\hat{V}_{e_{yx}e_{yx},\I}+\hat{V}_{e_{yc}e_{yc},\II}.$

We can construct the asymptotic $(1-\alpha)$ confidence interval
for $\bar{y}_{0}$ based on $\bar{y}_{\III}$ as
\[
\left(\bar{y}_{\III}-n_{\III}^{-1/2}\nu_{1-\alpha/2}(\hat{V}_{\III,1},\cdots,\hat{V}_{\III,5}),\ \bar{y}_{\III}-n_{\III}^{-1/2}\nu_{\alpha/2}(\hat{V}_{\III,1},\cdots,\hat{V}_{\III,5})\right),
\]
where $\nu_{\alpha}(V_{\III,1},\cdots,V_{\III,5})$ as the $100\alpha$th
quantile of the distribution of $V_{\III,1}^{1/2}L_{p_{1}+p_{2},\gamma_{2}^{2}}+V_{\III,2}^{1/2}Z_{1}+V_{\III,3}^{1/2}L_{p_{1},\gamma_{1}^{2}}+V_{\III,4}^{1/2}Z_{2}+V_{\III,5}^{1/2}Z_{3}$,
and the counterpart based on $\bar{y}_{\II,\reg}$ as
\[
\left(\bar{y}_{\III,\reg}-\hat{V}_{\III,\reg}^{1/2}z_{1-\alpha/2},\ \bar{y}_{\III}-\hat{V}_{\III,\reg}^{1/2}z_{\alpha/2}\right).
\]

\section{Proofs \protect\label{sec:Proofs}}

\subsection{Useful lemmas}

We state some useful lemmas for two-phase simple random sampling.

\begin{lemma}\label{lemma=000020A0} $E(\bar{u}_{\II}\mid\cA,\F)=\bar{u}_{\I}$
and $E(\bar{u}_{\I}\mid\F)=\bar{u}_{0}$.

\end{lemma}

\begin{lemma}\label{lemma=000020A1}$\cov(\bar{u}_{\II},\bar{v}_{\II}\mid\cA,\F)=\left(n_{\II}^{-1}-n_{\I}^{-1}\right)V_{uv,\I}$
and $\cov(\bar{u}_{\I},\bar{v}_{\I}\mid\F)=\left(n_{\I}^{-1}-N^{-1}\right)V_{uv,0}$.

\end{lemma}

\begin{lemma}\label{lemma=000020A2}$E(V_{uv,\II}\mid\cA,\F)=V_{uv,\I}$
and $E(V_{uv,\I}\mid\F)=V_{uv,0}$.

\end{lemma}

Lemmas \ref{lemma=000020A1} and \ref{lemma=000020A2} are standard
textbook results \citep[e.g.,][]{fuller2009sampling}; therefore,
we omit their proofs.

For a sequence of finite populations, the variability of the estimators
comes from the sampling design. In the following proofs for the asymptotic
design properties, the asymptotic design variance and covariance (the
limits of the design variance and covariances), denoted by \textcolor{black}{$\avar$
and $\acov$, respectively, are relevant. }

\subsection{Proof of Lemma \ref{lemma1_srs}}

Recall the following definitions $T_{1}=n_{\II}^{1/2}(\bar{x}_{\II}-\bar{x}_{\I})^{\T}\beta_{0}$,
$T_{2}=n_{\II}^{1/2}\left\{ \bar{y}_{\II}-\bar{y}_{\I}-(\bar{x}_{\II}-\bar{x}_{\I})^{\T}\beta_{0}\right\} $,
and $T_{3}=n_{\II}^{1/2}(\bar{y}_{\I}-\bar{y}_{0})$. The asymptotic
normality follows by Assumption \ref{asmp:sampling}; see, e.g., Chapter
1 in \citet{fuller2009sampling}.

Because $\cB$ is a simple random sample from $\cA$, and $\cA$ is
a simple random sample from $\F$, by Lemma \ref{lemma=000020A0},
we have 
\begin{equation}
E(T_{1}\mid\cA,\F)=0,\ E(T_{2}\mid\cA,\F)=0,\ E(T_{3}\mid\F)=0,\label{eq:A1}
\end{equation}
and therefore $E(T_{k}\mid\F)=0$, for $k=1,2,3$.

We then show the asymptotic variance formulas. First, the asymptotic
design variance of $T_{1}$ given $\F$ is 
\begin{eqnarray*}
\avar(T_{1}\mid\F_{N}) & \equiv & \lim_{N\rightarrow\infty}\var(T_{1}\mid\F)\\
 & = & \lim_{N\rightarrow\infty}\left[\var\left\{ E(T_{1}\mid\cA,\F)\mid\F\right\} +E\left\{ \var(T_{1}\mid\cA,\F)\mid\F\right\} \right]\\
 & = & \lim_{N\rightarrow\infty}\left[0+E\left\{ \left(1-\frac{n_{\II}}{n_{\I}}\right)\beta_{0}^{\T}V_{xx,\I}\beta_{0}\mid\F\right\} \right]\\
 & = & \lim_{N\rightarrow\infty}\left(1-\frac{n_{\II}}{n_{\I}}\right)\beta_{0}^{\T}V_{xx,0}\beta_{0}\\
 & = & \lim_{N\rightarrow\infty}\left(1-\frac{n_{\II}}{n_{\I}}\right)V_{yx,0}V_{xx,0}^{-1}V_{xy,0}\\
 & = & \left(1-f_{\II,\I}\right)\sigma_{yx}\sigma_{xx}^{-1}\sigma_{xy},\ \as,
\end{eqnarray*}
where the second equality follows by (\ref{eq:A1}) and Lemma \ref{lemma=000020A1},
and the third equality follows by Lemma \ref{lemma=000020A2}. Second,
by writing $T_{2}=n_{\II}^{1/2}(\bar{e}_{\II}-\bar{e}_{\I})+o_{P}(1)$,
the asymptotic design variance of $T_{2}$ given $\F$ is 
\begin{eqnarray*}
\avar(T_{2}\mid\F) & \equiv & \lim_{N\rightarrow\infty}\var(T_{2}\mid\F)\\
 & = & \lim_{N\rightarrow\infty}\left[\var\left\{ E(T_{2}\mid\cA,\F)\mid\F\right\} +E\left\{ \var(T_{2}\mid\cA,\F)\mid\F\right\} \right]\\
 & = & \lim_{N\rightarrow\infty}\left[0+E\left\{ \left(1-\frac{n_{\II}}{n_{\I}}\right)V_{ee,\I}\mid\F\right\} \right]\\
 & = & \lim_{N\rightarrow\infty}\left(1-\frac{n_{\II}}{n_{\I}}\right)V_{ee,0}\\
 & = & \left(1-f_{\II,\I}\right)\sigma_{ee},\ \as,
\end{eqnarray*}
where the second equality follows by (\ref{eq:A1}) and Lemma \ref{lemma=000020A1},
and the third equality follows by Lemma \ref{lemma=000020A2}. Third,
the asymptotic design variance of $T_{3}$ given $\F$ is $\avar(T_{3}\mid\F)\equiv\lim_{N\rightarrow\infty}\var(T_{3}\mid\F)=\lim_{N\rightarrow\infty}(n_{\II}/n_{\I}-n_{\II}/N)V_{yy,0}=f_{\II,\I}(1-f_{\I,0})\sigma_{yy}$,
$\as$.

Finally, we show that all covariances of cross terms are zero asymptotically.
We have 
\begin{eqnarray}
\acov(T_{1},T_{2}\mid\cA,\F) & \equiv & \lim_{N\rightarrow\infty}\cov(T_{1},T_{2}\mid\cA,\F)\nonumber \\
 & = & \lim_{N\rightarrow\infty}n_{\II}\beta_{0}^{\T}\cov(\bar{x}_{\II}-\bar{x}_{\I},\bar{y}_{\II}-\bar{y}_{\I}\mid\cA,\F)\nonumber \\
 &  & -\lim_{N\rightarrow\infty}n_{\II}\beta_{0}^{\T}\cov(\bar{x}_{\II}-\bar{x}_{\I},\bar{x}_{\II}-\bar{x}_{\I}\mid\cA,\F)\beta_{0}\nonumber \\
 & = & \lim_{N\rightarrow\infty}\left(1-\frac{n_{\II}}{n_{\I}}\right)\left(\beta_{0}^{\T}V_{xy,\I}-\beta_{0}^{\T}V_{xx,\I}\beta_{0}\right)\nonumber \\
 & = & \lim_{N\rightarrow\infty}\left(1-\frac{n_{\II}}{n_{\I}}\right)\left(V_{yx,0}V_{xx,0}^{-1}V_{xy,\I}-V_{yx,0}V_{xx,0}^{-1}V_{xx,\I}V_{xx,0}^{-1}V_{xy,0}\right)\nonumber \\
 & = & \left(1-f_{\II,\I}\right)\left(\sigma_{yx}\sigma_{xx}^{-1}\sigma_{xy}-\sigma_{yx}\sigma_{xx}\sigma_{xx}^{-1}\sigma_{xx}\sigma_{xy}\right)=0,\ \as,\label{eq:A2}
\end{eqnarray}
where the second equality follows by Lemma \ref{lemma=000020A2}.
Because conditional on $\cA$ and $\F$, $T_{3}$ is a constant, we
then have 
\begin{equation}
\acov(T_{k},T_{3}\mid\cA,\F)=0,\ k=1,2.\label{eq:A3}
\end{equation}
Combining (\ref{eq:A1})--(\ref{eq:A3}), we have 
\begin{eqnarray*}
\acov(T_{j},T_{k}\mid\F) & \equiv & \lim_{N\rightarrow\infty}\cov(T_{j},T_{k}\mid\F)\\
 & = & \lim_{N\rightarrow\infty}E\{\cov(T_{j},T_{k}\mid\cA,\F)\mid\F\}\\
 &  & +\lim_{N\rightarrow\infty}\cov\{E(T_{j}\mid\cA,\F),E(T_{k}\mid\cA,\F)\mid\F\}\\
 & = & 0,\ \left(j\neq k\in\{1,2,3\}\right),\ \as.
\end{eqnarray*}

Combining all results, we complete the proof.

\subsection{Proof of Theorem \ref{thm:2preg-rej}}

To express (\ref{eq:A4}) further, by the moment condition in Assumption
\ref{asmp:sampling}, we have $n_{\II}^{1/2}(\bar{x}_{\II}-\bar{x}_{\I})=O_{P}(1)$,
\begin{eqnarray*}
\hat{\beta}_{\II}-\beta_{0} & = & \left\{ \sum_{i\in\cB}\left(x_{i}-\bar{x}_{\II}\right)\left(x_{i}-\bar{x}_{\II}\right)^{\T}\right\} ^{-1}\sum_{i\in\cB}\left(x_{i}-\bar{x}_{\II}\right)\left\{ y_{i}-\bar{y}_{\II}-\left(x_{i}-\bar{x}_{\II}\right)^{\T}\beta_{0}\right\} \\
 & = & \left\{ n_{\II}^{-1}\sum_{i\in\cB}\left(x_{i}-\bar{x}_{\II}\right)\left(x_{i}-\bar{x}_{\II}\right)^{\T}\right\} ^{-1}n_{\II}^{-1}\sum_{i\in\cB}\left(x_{i}-\bar{x}_{\II}\right)\left(e_{i}-\bar{e}_{\II}\right)\\
 & = & O_{P}(n_{\II}^{-1/2}),
\end{eqnarray*}
and therefore $n_{\II}^{1/2}(\bar{x}_{\II}-\bar{x}_{\I})^{\T}(\hat{\beta}_{\II}-\beta_{0})=O_{P}(1)O_{P}(n_{\II}^{-1/2})=O_{P}(n_{\II}^{-1/2})=o_{P}(1).$
We then express (\ref{eq:A4}) as 
\begin{equation}
n_{\II}^{1/2}\left(\bar{y}_{\II,\reg}-\bar{y}_{0}\right)=n_{\II}^{1/2}\left(\bar{e}_{\II}-\bar{e}_{\I}\right)+n_{\II}^{1/2}\left(\bar{y}_{\I}-\bar{y}_{0}\right)+o_{P}(1).\label{eq:A5}
\end{equation}

Under TPRS in Definition \ref{def:TPRS}, based on (\ref{eq:A5}),
the design variance $\var\{n_{\II}^{1/2}(\bar{y}_{\II,\reg}-\bar{y}_{0})\mid\F\}$
is approximately 
\begin{eqnarray*}
n_{\II}\var\left(\bar{y}_{\I}-\bar{y}_{0}\mid\F\right)+n_{\II}E\left\{ \var(\bar{e}_{\II}\mid\cA,\F)\mid\F\right\}  & = & n_{\II}\left(\frac{1}{n_{\I}}-\frac{1}{N}\right)V_{yy,0}+E\left\{ \left(1-\frac{n_{\II}}{n_{\I}}\right)V_{ee,\I}\mid\F\right\} \\
 & = & n_{\II}\left(\frac{1}{n_{\I}}-\frac{1}{N}\right)V_{yy,0}+\left(1-\frac{n_{\II}}{n_{\I}}\right)V_{ee,0}\\
 & \rightarrow & f_{\II,\I}\left(1-f_{\I,0}\right)\sigma_{yy}+\left(1-f_{\II,\I}\right)\sigma_{ee},\ \as,
\end{eqnarray*}
where the second equality follows by Lemma \ref{lemma=000020A2}.

Under TPRS in Definition \ref{def:TPRS}, $n_{\II}^{1/2}(\bar{y}_{\II,\reg}-\bar{y}_{0})$
is equivalent to $n_{\II}^{1/2}(\bar{y}_{\II,\reg}-\bar{y}_{0})\mid(Q_{\I}<\gamma^{2})$
without rejective sampling. Similar to (\ref{eq:A2}), we can show
that $\bar{e}_{\II}-\bar{e}_{\I}$ and $Q_{\I}$ are asymptotically
independent. Moreover, we can show that $\bar{y}_{\I}-\bar{y}_{0}$
and $Q_{\I}$ are asymptotically independent. Based on (\ref{eq:A5}),
$n_{\II}^{1/2}(\bar{y}_{\II,\reg}-\bar{y}_{0})$ and $Q_{\I}$ are
asymptotically independent. As a result, $n_{\II}^{1/2}(\bar{y}_{\II,\reg}-\bar{y}_{0})$
has the same distribution under two-phase sampling with and without
rejective sampling.

\subsection{Proof of Proposition \ref{prop:ve}}

We follow the proofs in \citet{li2017general}. Under TPRS in Definition
\ref{def:TPRS}, $\bar{u}_{\II}\rightarrow E(u)$ $\as$, and 
\begin{eqnarray*}
\hat{V}_{uv} & = & \frac{1}{n_{\II}-1}\sum_{i\in\cB}(u_{i}-\bar{u}_{\II})(v_{i}-\bar{v}_{\II})\\
 & = & \frac{1}{n_{\II}-1}\sum_{i\in\cB}\{u_{i}-E(u)\}\{v_{i}-E(v)\}-\frac{n_{\II}}{n_{\II}-1}\{\bar{u}_{\II}-E(u)\}\{\bar{v}_{\II}-E(v)\}\\
 & \rightarrow & V_{uv},\ \as,
\end{eqnarray*}
where the last line follows from Assumption \ref{asmp:sampling}.
Next, we have 
\begin{eqnarray}
E\left\{ (\hat{V}_{uv}-V_{uv})^{2}\right\}  & = & \left(\frac{n_{\II}}{n_{\II}-1}\right)^{2}\var\left[\frac{1}{n_{\II}}\sum_{i\in\cB}\{u_{i}-E(u)\}\{v_{i}-E(v)\}-\{\bar{u}_{\II}-E(u)\}\{\bar{v}_{\II}-E(v)\}\right]\nonumber \\
 & = & \left(\frac{n_{\II}}{n_{\II}-1}\right)^{2}\var\left[\frac{1}{n_{\II}}\sum_{i\in\cB}\{u_{i}-E(u)\}\{v_{i}-E(v)\}\right]\nonumber \\
 &  & +\var\left[\{\bar{u}_{\II}-E(u)\}\{\bar{v}_{\II}-E(v)\}\right].\label{eq:hat-V}
\end{eqnarray}
To further evaluate (\ref{eq:hat-V}), we have 
\begin{eqnarray*}
\var\left[\frac{1}{n_{\II}}\sum_{i\in\cB}\{u_{i}-E(u)\}\{v_{i}-E(v)\}\right] & = & \left(\frac{1}{n_{\II}}\right)^{2}E\left(\sum_{i\in\cB}\left[\vphantom{\sum_{j\in\cB}}\{u_{i}-E(u)\}\{v_{i}-E(v)\}\right.\right.\\
 &  & -\left.\left.\frac{1}{n_{\II}}\sum_{j\in\cB}\{u_{j}-E(u)\}\{v_{j}-E(v)\}\right]^{2}\right)\\
 & \leq & \left(\frac{1}{n_{\II}}\right)^{2}E\left[\sum_{i\in\cB}\{u_{i}-E(u)\}^{2}\{v_{i}-E(v)\}^{2}\right]\\
 & \leq & \frac{1}{n_{\II}}E\left[\frac{1}{n_{\II}}\sum_{i\in\cB}\{u_{i}-E(u)\}^{4}+\frac{1}{n_{\II}}\sum_{i\in\cB}\{v_{i}-E(v)\}^{4}\right]\\
 & = & \frac{1}{n_{\II}}E\left[\{u-E(u)\}^{4}\right]+\frac{1}{n_{\II}}E\left[\{v-E(v)\}^{4}\right]
\end{eqnarray*}
and 
\begin{eqnarray*}
\var\left[\{\bar{u}_{\II}-E(u)\}\{\bar{v}_{\II}-E(v)\}\right] & \leq & E\left[\{\bar{u}_{\II}-E(u)\}^{2}\{\bar{v}_{\II}-E(v)\}^{2}\right]\\
 & \leq & E\left[\{\bar{u}_{\II}-E(u)\}^{4}+\{\bar{v}_{\II}-E(v)\}^{4}\right]\\
 & = & \frac{1}{n_{\II}}E\left[\{u-E(u)\}^{4}\right]+\frac{1}{n_{\II}}E\left[\{v-E(v)\}^{4}\right].
\end{eqnarray*}
Therefore, under Assumption \ref{asmp:sampling}, 
\begin{equation}
E\left\{ (\hat{V}_{uv}-V_{uv})^{2}\right\} \leq\frac{2}{n_{\II}}E\left[\{u-E(u)\}^{4}\right]+\frac{2}{n_{\II}}E\left[\{v-E(v)\}^{4}\right]\rightarrow0.\label{eq:hat-V2}
\end{equation}

By the law of total probability, we have
\begin{eqnarray*}
E\left\{ (\hat{V}_{uv}-V_{uv})^{2}\right\}  & = & E\left\{ (\hat{V}_{uv}-V_{uv})^{2}\mid Q_{\I}<\gamma^{2}\right\} \pr(Q_{\I}<\gamma^{2})\\
 &  & +E\left\{ (\hat{V}_{uv}-V_{uv})^{2}\mid Q_{\I}\geq\gamma^{2}\right\} \pr(Q_{\I}\geq\gamma^{2})\\
 & \geq & E\left\{ (\hat{V}_{uv}-V_{uv})^{2}\mid Q_{\I}<\gamma^{2}\right\} \pr(Q_{\I}<\gamma^{2}).
\end{eqnarray*}
Combining the above inequality with (\ref{eq:hat-V2}), we obtain
\begin{eqnarray*}
E\left\{ (\hat{V}_{uv}-V_{uv})^{2}\mid Q_{\I}<\gamma^{2}\right\}  & \leq & \{\pr(Q_{\I}<\gamma^{2})\}^{-1}E\left\{ (\hat{V}_{uv}-V_{uv})^{2}\right\} =o(1).
\end{eqnarray*}
Therefore, $\hat{V}_{uv}=V_{uv}+o_{P}(1)$, which completes the proof.

\subsection{Proof of Lemma \ref{Lemma2-pi}}

By the constructions of phase $\I$ and $\II$ samples and estimators,
we have 
\begin{equation}
E(T_{1}\mid\cA,\F)=0,\ E(T_{2}\mid\cA,\F)=0,\ E(T_{3}\mid\F)=0,\label{eq:A1-1}
\end{equation}
and therefore $E(T_{k}\mid\F)=0$, for $k=1,2,3$.

We then show the asymptotic variance formulas. First, the asymptotic
design variance of $T_{1}$ given $\F$ is 
\begin{multline*}
\avar(T_{1}\mid\F)\equiv\lim_{N\rightarrow\infty}\var(T_{1}\mid\F)=\lim_{N\rightarrow\infty}\left[\var\left\{ E(T_{1}\mid\cA,\F)\mid\F\right\} +E\left\{ \var(T_{1}\mid\cA,\F)\mid\F\right\} \right]\\
=\lim_{N\rightarrow\infty}\left[0+E\left\{ \frac{n_{\II}}{N^{2}}\sum_{i\in\cA}\sum_{j\in\cA}\frac{\pi_{\II ij\mid\cA}-\pi_{\II i\mid\cA}\pi_{\II j\mid\cA}}{\pi_{i}^{*}\pi_{j}^{*}}(y_{i}-\bar{y}_{0})(y_{j}-\bar{y}_{0})\mid\F\right\} \right]=V_{1},\ \as.
\end{multline*}
Second, the asymptotic design variance of $T_{2}$ given $\F$ is
\begin{multline*}
\avar(T_{2}\mid\F)\equiv\lim_{N\rightarrow\infty}\var(T_{2}\mid\F)=\lim_{N\rightarrow\infty}\left[\var\left\{ E(T_{2}\mid\cA,\F)\mid\F\right\} +E\left\{ \var(T_{2}\mid\cA,\F)\mid\F\right\} \right]\\
=\lim_{N\rightarrow\infty}\left[0+E\left\{ \frac{n_{\II}}{N^{2}}\sum_{i\in\cA}\sum_{j\in\cA}\frac{\pi_{\II ij\mid\cA}-\pi_{\II i\mid\cA}\pi_{\II j\mid\cA}}{\pi_{i}^{*}\pi_{j}^{*}}(e_{i}-\bar{e}_{0})(e_{j}-\bar{e}_{0})\mid\F\right\} \right]=V_{2},\ \as.
\end{multline*}
Third, the asymptotic design variance of $T_{3}$ given $\F$ is 
\begin{multline*}
\avar(T_{3}\mid\F)\equiv\lim_{N\rightarrow\infty}\var(T_{3}\mid\F)\\
=\lim_{N\rightarrow\infty}\left\{ \frac{n_{\II}}{N^{2}}\sum_{i=1}^{N}\sum_{j=1}^{N}\frac{\pi_{\I ij}-\pi_{\I i}\pi_{\I j}}{\pi_{\I i}\pi_{\I j}}(y_{i}-\bar{y}_{0})(y_{j}-\bar{y}_{0})\right\} =V_{3},\ \as.
\end{multline*}
Note that $V_{1}$, $V_{2}$ and $V_{3}$ are finite, guaranteed by
Assumption \ref{asmp:sampling-pi}.

Finally, we show that all covariances of cross terms are zero asymptotically.
By Assumption \ref{asmp:sampling-pi}(vii), $\acov(T_{1},T_{2}\mid\F)=0$
$\as$. Because conditional on $\cA$, $T_{3}$ is a constant, we
then have 
\begin{equation}
\acov(T_{k},T_{3}\mid\cA,\F)=0,\ k=1,2.\label{eq:A3-1}
\end{equation}
Combining (\ref{eq:A1-1})--(\ref{eq:A3-1}), we have 
\begin{eqnarray*}
\acov(T_{j},T_{k}\mid\F) & \equiv & \lim_{N\rightarrow\infty}\cov(T_{j},T_{k}\mid\F)\\
 & = & \lim_{N\rightarrow\infty}E\{\cov(T_{j},T_{k}\mid\cA,\F)\mid\F\}\\
 &  & +\lim_{N\rightarrow\infty}\cov\{E(T_{j}\mid\cA,\F),E(T_{k}\mid\cA,\F)\mid\F\}\\
 & = & 0,\ \left(j\neq k\in\{1,2,3\}\right),\ \as.
\end{eqnarray*}

Combining all results completes the proof.

\subsection{Proof of Theorem \ref{Thm:reg-pi}}

To derive the asymptotic design property of the two-phase regression
estimator, we use the following decomposition:
\begin{eqnarray*}
\bar{y}_{\II,\reg}-\bar{y}_{0} & = & \bar{y}_{\II}-\bar{y}_{\I}+(\bar{x}_{\I}-\bar{x}_{\II})^{\T}\beta_{0}+(\bar{x}_{\I}-\bar{x}_{\II})^{\T}(\hat{\beta}_{\II}-\beta_{0})+\bar{y}_{\I}-\bar{y}_{0}\\
 & = & \bar{e}_{\II}-\bar{e}_{\I}+\bar{y}_{\I}-\bar{y}_{0}+O_{P}(n_{\II}^{-1}),
\end{eqnarray*}
where the second equality follows by $(\bar{x}_{\I}-\bar{x}_{\II})^{\T}(\hat{\beta}_{\II}-\beta_{0})=O_{P}(n_{\II}^{-1})$
due to $\bar{x}_{\I}-\bar{x}_{\II}=O_{P}(n_{\II}^{-1/2})$ and $\hat{\beta}_{\II}-\beta_{0}=O_{P}(n_{\II}^{-1/2})$.

Therefore, the asymptotic design variance of $\bar{y}_{\II,\reg}$
is 
\begin{eqnarray*}
\avar\left\{ n_{\II}^{1/2}(\bar{y}_{\II,\reg}-\bar{y}_{0})\mid\F\right\}  & \equiv & \lim_{N\rightarrow\infty}\var\left\{ n_{\II}^{1/2}(\bar{y}_{\II,\reg}-\bar{y}_{0})\mid\F\right\} \\
 & = & \lim_{N\rightarrow\infty}\left[\var(n_{\II}^{1/2}\bar{y}_{\I}\mid\F)+E\left\{ \var(n_{\II}^{1/2}\bar{e}_{\II}\mid\cA,\F)\mid\F_{N}\right\} \right]\\
 & = & \lim_{N\rightarrow\infty}\left[\frac{n_{\II}}{N^{2}}\sum_{i=1}^{N}\sum_{j=1}^{N}\frac{\pi_{\I ij}-\pi_{\I i}\pi_{\I j}}{\pi_{\I i}\pi_{\I j}}(y_{i}-\bar{y}_{0})(y_{j}-\bar{y}_{0})\right.\\
 &  & +\left.E\left\{ \frac{n_{\II}}{N^{2}}\sum_{i\in\cA}\sum_{j\in\cA}\frac{\pi_{\II ij\mid\cA}-\pi_{\II i\mid\cA}\pi_{\II j\mid\cA}}{\pi_{i}^{*}\pi_{j}^{*}}(e_{i}-\bar{e}_{0})(e_{j}-\bar{e}_{0})\mid\F\right\} \right]\\
 & = & \lim_{N\rightarrow\infty}\left\{ n_{\II}V_{yy,0}+E\left(n_{\II}V_{ee,\I}\mid\F\right)\right\} .
\end{eqnarray*}

\subsection{Proof of Remark \ref{Remark:SYG}}

For the phase-$\I$ and phase-$\II$ sample sizes, we have 
\begin{eqnarray}
\sum_{j=1}^{N}I(j\in\cA) & = & n_{\I},\label{eq:s1}\\
\sum_{j\in\cA}I(j\in\cB) & = & n_{\II}.\label{eq:s2}
\end{eqnarray}
By taking the expectation of (\ref{eq:s1}), we obtain 
\begin{equation}
\sum_{j=1}^{N}\pi_{\I j}=n_{\I},\label{eq:s1-1}
\end{equation}
By taking the conditional expectation of (\ref{eq:s2}) given $\cA$,
we obtain 
\begin{equation}
\sum_{j\in\cA}\pi_{\II j\mid\cA}=n_{\II}.\label{eq:s2-1}
\end{equation}

Multiplying (\ref{eq:s1}) by $I(i\in\cA)$ and multiplying (\ref{eq:s2})
by $I(i\in\cB),$ we obtain 
\begin{eqnarray}
\sum_{j=1}^{N}I(j\in\cA)I(i\in\cA) & = & n_{\I}I(i\in\cA),\label{eq:s1-2}\\
\sum_{j\in\cA}I(j\in\cB)I(i\in\cB) & = & n_{\II}I(i\in\cB).\label{eq:s2-2}
\end{eqnarray}
By taking the expectation of (\ref{eq:s1-2}), we obtain 
\begin{equation}
\sum_{j=1}^{N}\pi_{\I ij}=n_{\I}\pi_{\I i},\label{eq:s1-3}
\end{equation}
By taking the conditional expectation of (\ref{eq:s2-2}) given $\cA$,
we obtain 
\begin{equation}
\sum_{j\in\cA}\pi_{\II ij\mid\cA}=n_{\II}\pi_{\II i\mid\cA}.\label{eq:s2-3}
\end{equation}

We now write 
\begin{eqnarray*}
\hat{V}_{2,\mathrm{SYG}} & = & -\frac{n_{\II}}{N^{2}}\sum_{i\in\cB}\sum_{j\in\cB}\frac{\pi_{\II ij\mid\cA}-\pi_{\II i\mid\cA}\pi_{\II j\mid\cA}}{\pi_{\II ij\mid\cA}}\left(\frac{\hat{e}_{i}-\hat{e}_{\II}}{\pi_{\II i}^{*}}\right)^{\otimes2}\\
 &  & +\frac{n_{\II}}{N^{2}}\sum_{i\in\cB}\sum_{j\in\cB}\frac{\pi_{\II ij\mid\cA}-\pi_{\II i\mid\cA}\pi_{\II j\mid\cA}}{\pi_{\II ij\mid\cA}}\left(\frac{\hat{e}_{i}-\hat{e}_{\II}}{\pi_{\II i}^{*}}\right)\left(\frac{\hat{e}_{j}-\hat{e}_{\II}}{\pi_{\II j}^{*}}\right)^{\T}\\
 & = & T_{2,\mathrm{SYG}}+\hat{V}_{2}.
\end{eqnarray*}
Moreover,
\begin{eqnarray}
E\left(T_{2,\mathrm{SYG}}\mid\cA,\F\right) & = & -\frac{n_{\II}}{N^{2}}E\left\{ \sum_{i\in\cA}\sum_{j\in\cA}\left(\pi_{\II ij\mid\cA}-\pi_{\II i\mid\cA}\pi_{\II j\mid\cA}\right)\left(\frac{\hat{e}_{i}-\hat{e}_{\II}}{\pi_{\II i}^{*}}\right)^{\otimes2}\mid\cA,\F\right\} \nonumber \\
 & = & -\frac{n_{\II}}{N^{2}}E\left\{ \sum_{i\in\cA}\left(\sum_{j\in\cA}\pi_{\II ij\mid\cA}\right)\left(\frac{\hat{e}_{i}-\hat{e}_{\II}}{\pi_{\II i}^{*}}\right)^{\otimes2}\mid\cA,\F\right\} \nonumber \\
 &  & +\frac{n_{\II}}{N^{2}}E\left\{ \sum_{i\in\cA}\pi_{\II i\mid\cA}\left(\sum_{j\in\cA}\pi_{\II j\mid\cA}\right)\left(\frac{\hat{e}_{i}-\hat{e}_{\II}}{\pi_{\II i}^{*}}\right)^{\otimes2}\mid\cA,\F\right\} \nonumber \\
 & = & -\frac{n_{\II}}{N^{2}}E\left\{ \sum_{i\in\cA}\left(n_{\II}\pi_{\II i\mid\cA}\right)\left(\frac{\hat{e}_{i}-\hat{e}_{\II}}{\pi_{\II i}^{*}}\right)^{\otimes2}\mid\cA,\F\right\} \label{eq:s2-implied1}\\
 &  & +\frac{n_{\II}}{N^{2}}E\left\{ \sum_{i\in\cA}\pi_{\II i\mid\cA}n_{\II}\left(\frac{\hat{e}_{i}-\hat{e}_{\II}}{\pi_{\II i}^{*}}\right)^{\otimes2}\mid\cA,\F\right\} \label{eq:s2-implied2}\\
 & = & 0,\nonumber 
\end{eqnarray}
where (\ref{eq:s2-implied1}) follows by (\ref{eq:s2-3}), and (\ref{eq:s2-implied2})
follows by (\ref{eq:s2-1}). Therefore, $\hat{V}_{2,\mathrm{SYG}}$
is asymptotically equivalent to $\hat{V}_{2}$ adding a mean zero
term.

We also write 
\begin{eqnarray*}
\hat{V}_{3,\mathrm{SYG}} & = & -\frac{n_{\II}}{N^{2}}\sum_{i\in\cB}\sum_{j\in\cB}\left(\frac{\pi_{\I ij}-\pi_{\I i}\pi_{\I j}}{\pi_{\I ij}\pi_{\II ij\mid\cA}}\right)\left(\frac{y_{i}-\bar{y}_{\II}}{\pi_{\I i}}\right)^{\otimes2}\\
 &  & +\frac{n_{\II}}{N^{2}}\sum_{i\in\cB}\sum_{j\in\cB}\left(\frac{\pi_{\I ij}-\pi_{\I i}\pi_{\I j}}{\pi_{\I ij}\pi_{\II ij\mid\cA}}\right)\left(\frac{y_{i}-\bar{y}_{\II}}{\pi_{\I i}}\right)\left(\frac{y_{j}-\bar{y}_{\II}}{\pi_{\I j}}\right)^{\T}\\
 & = & T_{3,\mathrm{SYG}}+\hat{V}_{3}.
\end{eqnarray*}
Moreover, 
\begin{eqnarray}
E(T_{3,\mathrm{SYG}}) & = & -\frac{n_{\II}}{N^{2}}E\left\{ \sum_{i=1}^{N}\left(\sum_{j=1}^{N}\pi_{\I ij}\right)\left(\frac{y_{i}-\bar{y}_{\II}}{\pi_{\I i}}\right)^{\otimes2}-\left(\sum_{j=1}^{N}\pi_{\I j}\right)\sum_{i=1}^{N}\pi_{\I i}\left(\frac{y_{i}-\bar{y}_{\II}}{\pi_{\I i}}\right)^{\otimes2}\right\} \nonumber \\
 & = & -\frac{n_{\II}}{N^{2}}E\left\{ \sum_{i=1}^{N}\left(\sum_{j=1}^{N}\pi_{\I ij}\right)\left(\frac{y_{i}-\bar{y}_{\II}}{\pi_{\I i}}\right)^{\otimes2}-\left(\sum_{j=1}^{N}\pi_{\I j}\right)\sum_{i=1}^{N}\pi_{\I i}\left(\frac{y_{i}-\bar{y}_{\II}}{\pi_{\I i}}\right)^{\otimes2}\right\} \nonumber \\
 & = & -\frac{n_{\II}}{N^{2}}E\left\{ \sum_{i=1}^{N}n_{\I}\pi_{\I i}\left(\frac{y_{i}-\bar{y}_{\II}}{\pi_{\I i}}\right)^{\otimes2}-n_{\I}\sum_{i=1}^{N}\pi_{\I i}\left(\frac{y_{i}-\bar{y}_{\II}}{\pi_{\I i}}\right)^{\otimes2}\right\} =0,\label{eq:s3-implied}
\end{eqnarray}
where (\ref{eq:s3-implied}) follows by (\ref{eq:s1-1}) and (\ref{eq:s1-3}).
Therefore, $\hat{V}_{3,\mathrm{SYG}}$ is asymptotically equivalent
to $\hat{V}_{3}$ adding a mean zero term.

\subsection{Proof Theorem \ref{Thm:3}}

We use the following decomposition: 
\begin{eqnarray*}
n_{\III}^{1/2}\left(\bar{y}_{\III}-\bar{y}_{0}\right) & = & n_{\III}^{1/2}(\bar{y}_{\III}-\bar{y}_{\II})+n_{\III}^{1/2}(\bar{y}_{\II}-\bar{y}_{\I})+n_{\III}^{1/2}(\bar{y}_{\I}-\bar{y}_{0})\\
 & = & n_{\III}^{1/2}(\bar{c}_{\III}-\bar{c}_{\II})\beta_{yc,0}+n_{\III}^{1/2}\left\{ \bar{y}_{\III}-\bar{y}_{\II}-(\bar{c}_{\III}-\bar{c}_{\II})\beta_{yc,0}\right\} \\
 &  & +n_{\III}^{1/2}(\bar{x}_{\II}-\bar{x}_{\I})\beta_{yx,0}+n_{\III}^{1/2}\left\{ \bar{y}_{\II}-\bar{y}_{\I}-(\bar{x}_{\II}-\bar{x}_{\I})\beta_{yx,0}\right\} \\
 &  & +n_{\III}^{1/2}(\bar{y}_{\I}-\bar{y}_{0})\\
 & = & n_{\III}^{1/2}(\bar{c}_{\III}-\bar{c}_{\II})\beta_{yc,0}+n_{\III}^{1/2}\left(\bar{e}_{yc,\III}-\bar{e}_{yc,\II}\right)\\
 &  & +n_{\III}^{1/2}(\bar{x}_{\II}-\bar{x}_{\I})\beta_{yx,0}+n_{\III}^{1/2}\left(\bar{e}_{yx,\II}-\bar{e}_{yx,\I}\right)\\
 &  & +n_{\III}^{1/2}(\bar{y}_{\I}-\bar{y}_{0}).
\end{eqnarray*}
We have $D_{x,\I}=V_{xx,\I}^{-1/2}\left(\bar{x}_{\II}-\bar{x}_{\I}\right)\rightarrow\N(0,I_{p_{1}})$
and $D_{c,\II}=V_{cc,\II}^{-1/2}\left(\bar{c}_{\III}-\bar{c}_{\II}\right)\rightarrow\N(0,I_{p_{1}+p_{2}})$.
Therefore, we have 
\begin{eqnarray*}
 &  & n_{\III}^{1/2}\left(\bar{y}_{\III}-\bar{y}_{0}\right)\mid(Q_{x,\I}<\gamma_{1}^{2},Q_{c,\II}<\gamma_{2}^{2})\\
 & = & n_{\III}^{1/2}(\bar{c}_{\III}-\bar{c}_{\II})\beta_{yc,0}\mid(Q_{x,\I}<\gamma_{1}^{2},Q_{c,\II}<\gamma_{2}^{2})+n_{\III}^{1/2}\left(\bar{e}_{yc,\III}-\bar{e}_{yc,\II}\right)\\
 &  & +n_{\III}^{1/2}(\bar{x}_{\II}-\bar{x}_{\I})\beta_{yx,0}\mid(Q_{x,\I}<\gamma_{1}^{2},Q_{c,\II}<\gamma_{2}^{2})+n_{\III}^{1/2}\left(\bar{e}_{yx,\II}-\bar{e}_{yx,\I}\right)\\
 &  & +n_{\III}^{1/2}(\bar{y}_{\I}-\bar{y}_{0})\\
 & = & n_{\III}^{1/2}(\bar{c}_{\III}-\bar{c}_{\II})\beta_{yc,0}\mid(Q_{c,\II}<\gamma_{2}^{2})+n_{\III}^{1/2}\left(\bar{e}_{yc,\III}-\bar{e}_{yc,\II}\right)\\
 &  & +n_{\III}^{1/2}(\bar{x}_{\II}-\bar{x}_{\I})\beta_{yx,0}\mid(Q_{x,\I}<\gamma_{1}^{2})+n_{\III}^{1/2}\left(\bar{e}_{yx,\II}-\bar{e}_{yx,\I}\right)\\
 &  & +n_{\III}^{1/2}(\bar{y}_{\I}-\bar{y}_{0})\\
 & \rightarrow & V_{\III,1}^{1/2}D_{c,\II}\mid\left(D_{c,\II}^{\T}D_{c,\II}<\gamma_{2}^{2}\right)+V_{\III,2}^{1/2}Z_{1}+V_{\III,3}^{1/2}D_{x,\I}\mid\left(D_{x,\I}^{\T}D_{x,\I}<\gamma_{1}^{2}\right)+V_{\III,4}^{1/2}Z_{2}+V_{\III,5}^{1/2}Z_{3}\\
 & \rightarrow & V_{\III,1}^{1/2}L_{p_{1}+p_{2},\gamma_{2}^{2}}+V_{\III,2}^{1/2}Z_{1}+V_{\III,3}^{1/2}L_{p_{1},\gamma_{1}^{2}}+V_{\III,4}^{1/2}Z_{2}+V_{\III,5}^{1/2}Z_{3}.
\end{eqnarray*}

\subsection{Proof of Theorem \ref{Thm3-reg}}

To derive the asymptotic design properties of the three-phase regression
estimator, we use the following facts: 
\[
E\left(\bar{y}_{\III,\reg}\mid\cB,\cA,\F\right)=\bar{y}_{\II,\reg}+O_{P}(n_{\III}^{-1}),
\]
where the probability distribution in $O_{P}$ is due to phase-III
random sampling given $(\cB,\cA,\F)$, $\bar{y}_{\II,\reg}=\bar{y}_{\II}+\left(\bar{x}_{\I}-\bar{x}_{\II}\right)^{\T}\hat{\beta}_{yx,\II}$,
and $\hat{\beta}_{yx,\II}$ is defined as (\ref{eq:hat-betaII}) with
$z$ and $x$ being $y$ and $x$, and 
\[
E\left(\bar{y}_{\II,\reg}\mid\cA,\F\right)=\bar{y}_{\I}+O_{P}(n_{\II}^{-1}),
\]
the probability distribution in $O_{P}$ is due to phase-II random
sampling given $(\cA,\F)$. Therefore, we use the following decomposition:
\begin{eqnarray*}
\bar{y}_{\III,\reg}-\bar{y}_{0} & = & \bar{y}_{\III,\reg}-\bar{y}_{\II,\reg}+\bar{y}_{\II,\reg}-\bar{y}_{\I}+\bar{y}_{\I}-\bar{y}_{0}\\
 & = & \left\{ \bar{y}_{\III}+\left(\begin{array}{c}
\bar{x}_{\I}-\bar{x}_{\III}\\
-\bar{a}_{\III}
\end{array}\right)^{\T}\hat{\beta}_{yc,\III}\right\} -\left\{ \bar{y}_{\II}+\left(\bar{x}_{\I}-\bar{x}_{\II}\right)^{\T}\hat{\beta}_{yx,\II}\right\} \\
 &  & +\left\{ \bar{y}_{\II}-\bar{y}_{\I}+\left(\bar{x}_{\I}-\bar{x}_{\II}\right)^{\T}\hat{\beta}_{yx,\II}\right\} +\bar{y}_{\I}-\bar{y}_{0}\\
 & = & \left(\bar{e}_{yc,\III}-\bar{e}_{yc,\II}\right)+\left(\bar{e}_{yx,\II}-\bar{e}_{yx,\I}\right)+\bar{y}_{\I}-\bar{y}_{0}+O_{P}(n_{\III}^{-1}).
\end{eqnarray*}
By repeated application of conditional expectation arguments, the
asymptotic design variance of $\bar{y}_{\III,\reg}$ is 
\[
\avar\left\{ n_{\III}^{1/2}(\bar{y}_{\III,\reg}-\bar{y}_{0})\mid\F\right\} =\lim_{N\rightarrow\infty}\left\{ n_{\III}V_{yy,0}+E\left(n_{\III}V_{e_{yx}e_{yx},\I}\mid\F\right)+E\left(n_{\III}V_{e_{yc}e_{yc},\II}\mid\F\right)\right\} .
\]
Then, under Assumptions \ref{asmp:sampling-pi} and \ref{asmp:sampling-pi-3phase},
we establish the result in Theorem \ref{Thm3-reg}.

\subsection{Proof of Theorem \ref{Thm-gen-genparameter}}

We first state the regularity conditions for Theorem \ref{Thm-gen-genparameter}.

\begin{assumption}\label{assumptionB:score} The following conditions
hold for the population parameter $\xi_{0}$ and the population estimating
function $\bar{s}_{0}(\cdot)$: 
\begin{enumerate}
\item[(i)]  The population parameter $\xi_{0}$ lies within a closed interval
$\mathcal{I}_{\xi}$.
\item[(ii)]  The function $s_{i}(\cdot)$ is bounded.
\item[(iii)]  The population estimating function $\bar{s}_{0}(\xi)$ converges
uniformly to $s_{0}(\xi)$ on $\mathcal{I}_{\xi}$ as $N\rightarrow\infty$,
and the equation $s_{0}(\xi)=0$ has a unique root within the interior
of $\mathcal{I}_{\xi}$.
\item[(iv)]  The limiting function $s_{0}(\xi)$ is strictly increasing in each
component of $\xi$ and absolutely continuous, with finite first and
second derivatives in $\mathcal{I}_{\xi}$. Additionally, the derivative
$\partial s_{0}(\xi)/\partial\xi$ is bounded away from zero within
$\mathcal{I}_{\xi}$.
\item[(v)]  The following conditions on the population quantities are satisfied:
\[
\sup_{\xi\in\mathcal{I}_{\xi}}N^{1/2}|\bar{s}_{0}(\xi_{0}+N^{-1/2}\xi)-\bar{s}_{0}(\xi_{0})-s_{0}(\xi_{0}+N^{-1/2}\xi)-s_{0}(\xi_{0})|\rightarrow0,
\]
and
\[
\sup_{\xi\in\mathcal{I}_{\xi}}N^{-1}\sum_{i=1}^{N}|s_{i}(\xi_{0}+N^{-1/2}\xi)-s_{i}(\xi_{0})|=O_{P}(N^{-1/2}).
\]
\end{enumerate}
\end{assumption}

Assumption \ref{assumptionB:score}(i)--(iv) are common for M-estimators
\citep{serfling1980approximation}. Assumption \ref{assumptionB:score}(v)
applies to differentiable estimating functions and has been examined
by \citet{wang2011asymptotic} for non-differentiable estimating functions.
\citet{wang2011asymptotic} demonstrate that under appropriate conditions
on the probability mechanism generating the $y_{i}$ values and the
function $s(y_{i};\xi)$, Assumption \ref{assumptionB:score}(v) holds
with probability one.

We will now show the results in Theorem \ref{Thm-gen-genparameter}.
Under Assumption \ref{assumptionB:score}, and using the theory for
M-estimators for differentiable estimating functions or the results
of \citet{wang2011asymptotic} for non-differentiable estimating functions,
we can express
\begin{equation}
\bar{s}_{\II}(\bar{\xi}_{\II})-\bar{s}_{0}(\xi_{0})=\bar{s}_{\II}(\xi_{0})-\bar{s}_{0}(\xi_{0})+\Gamma_{s}^{\T}(\bar{\xi}_{\II}-\xi_{0})+o_{P}(n_{\II}^{-1/2}).\label{eq:result-S}
\end{equation}
By Assumption \ref{assumptionB:score}(iv), $\bar{s}_{0}(\xi)$ is
smooth, implying $\bar{s}_{0}(\xi_{0})=O_{P}(N^{-1}),$ $\bar{s}_{\II}(\bar{\xi}_{\II})=O_{P}(n_{\II}^{-1}),$
and the left-hand side of (\ref{eq:result-S}) is $o_{P}(n_{\II}^{-1/2}).$ 

Because Assumption \ref{asmp:sampling-pi} holds for $s_{i}=s_{i}(\xi_{0})$,
we define the residual vector as 
\[
e_{i}^{s}=s_{i}-B_{0}^{\T}x_{i},
\]
where 
\[
B_{0}=\left\{ \sum_{i=1}^{N}(x_{i}-\bar{x}_{0})^{\otimes2}\right\} ^{-1}\sum_{i=1}^{N}(x_{i}-\bar{x}_{0})\{s_{i}(\xi_{0})-\bar{s}_{0}(\xi_{0})\}^{\T}.
\]
Under TPRS  in Definition \ref{def:TPRS_gen}, we have\textcolor{black}{
\begin{equation}
n_{\II}^{1/2}\{\bar{s}_{\II}(\xi_{0})-\bar{s}_{0}(\xi_{0})\}\mid\left(Q_{\I}<\gamma^{2}\right)\rightarrow(V_{1}^{s})^{1/2}L_{p,\gamma^{2}}+(V_{2}^{s})^{1/2}Z_{1}+(V_{3}^{s})^{1/2}Z_{2},\label{eq:result-S-2}
\end{equation}
}where $Z_{1}$ and $Z_{2}$ are standard normal variables, and $(L_{p,\gamma^{2}},Z_{1},Z_{2})$
are jointly independent.

Combining the results (\ref{eq:result-S}) and (\ref{eq:result-S-2}),
we obtain
\[
n_{\II}^{1/2}(\bar{\xi}_{\II}-\xi_{0})\mid\left(Q_{\I}<\gamma^{2}\right)\rightarrow\Gamma_{s}^{\T}(V_{1}^{s})^{1/2}L_{p,\gamma^{2}}+\Gamma_{s}^{\T}(V_{2}^{s})^{1/2}Z_{1}+\Gamma_{s}^{\T}(V_{3}^{s})^{1/2}Z_{2},
\]
as $n_{\II}\rightarrow\infty$. 
\end{document}